\documentclass[%
 reprint,
superscriptaddress,
 amsmath,amssymb,
 aps,
pra,
]{revtex4-1}

\usepackage{graphicx}
\usepackage{dcolumn}
\usepackage{bm}
\usepackage[T1]{fontenc}
\usepackage[utf8]{inputenc}
\usepackage{lmodern}
\usepackage{subfigure}
\usepackage{braket}

\newcommand{\pu}{\Phi_{\mathrm{u}}}
\newcommand{\pd}{\Phi_{\mathrm{d}}}
\newcommand{\p}{\Phi}

\newcommand{\geff}{g_{\textrm{eff}}}

\begin{document}

\preprint{APS/123-QED}

\title{Al'tshuler-Aronov-Spivak oscillations of bosonic matter-wave beams in the presence of interaction}

\newcommand{\LiegeUniversity}{CESAM research unit, University of Liège, 4000 Liège, Belgium}
\newcommand{\RegensburgUniversity}{Institut für Theoretische Physik, Universität Regensburg, 93040 Regensburg, Germany}

\author{Renaud Chr\'etien} \email{rchretien@ulg.ac.be} \affiliation{\LiegeUniversity}\email{rchretien@ulg.ac.be}
\author{Josef Rammensee} \affiliation{\RegensburgUniversity}
\author{Julien Dujardin} \affiliation{\LiegeUniversity}
\author{Cyril Petitjean} \affiliation{\LiegeUniversity}
\author{Peter Schlagheck} \affiliation{\LiegeUniversity}

\date{\today}

\begin{abstract}
We theoretically study the propagation of a guided atom laser across an Aharonov-Bohm ring which is exposed to a synthetic gauge field. The presence of disorder within the ring gives rise to Al'tshuler-Aronov-Spivak oscillations, seen in the disorder average of the transmission as a function of the effective gauge flux that is contained within the ring. Those oscillations are induced by coherent backscattering and represent a manifestation of weak localisation. Through analytical and numerical calculations that are based on the mean-field Gross-Pitaevskii approximation for the propagating Bose-Einstein condensate, we show that the presence of a very weak atom-atom interaction within the ring leads to an inversion of the Al'tshuler-Aronov-Spivak oscillations, in a very similar manner as for the coherent backscattering of Bose-Einstein condensates within two-dimensional disorder potentials. Numerical simulations based on the Truncated Wigner method reveal that this signature of weak antilocalisation becomes washed out if the interaction strength is increased.
\end{abstract}

\pacs{Valid PACS appear here}
\maketitle


\section{Introduction}
Weak localisation \cite{PhysRevB.22.5142,BERGMANN19841} is a physical effect related to a notable increase of the reflection of coherent waves traversing a disordered scattering region, compared with the incoherent transport process. As a result of current conservation, this increase is responsible for a drop in the transmission. This effect has been studied in mesoscopic physics for long now, as it highlights a macroscopic outcome resulting from quantum interferences. In solid state physics, positive corrections to the resistivity of a disordered sample, due to weak localisation, must indeed be added to classical predictions of the Drude formula \cite{Drude1900I,Drude1900II} describing electronic transport through a disordered sample \cite{akkermans_montambaux_2007}. They originate from a constructive interference between the contributions resulting from a scattering path and its time reversal counterpart, which survives the disorder average. 

A most prominent signature of quantum interference in mesoscopic physics is coherent backscattering \cite{WolfP.E.1988,refId0,MaretPRL1985}. This phenomenon, which is encountered in a wide variety of domains, is responsible for an enhancement of the backscattered current of a disordered sample that is illuminated by coherent waves, involving exactly the same mechanism as weak localisation. Coherent backscattering was already observed as soon as in 1893 to explain the fact that Saturn's ring are twice brighter in the backscattered direction \cite{HAPKE2002523}. More recentely, it was observed in laboratory by illuminating a powder with laser light \cite{PhysRevLett.55.2692,WolfP.E.1988}, but also for acoustic waves \cite{PhysRevLett.79.3637} and for elastic waves \cite{PhysRevLett.84.1693}. It is also used in seismology to probe the underground deeply, and for the research of oil \cite{Margerin2009}. 

More recently, coherent backscattering was also studied with matter waves by means of Bose-Einstein condensates \cite{PhysRevLett.109.195302}. In this context, new questions related to many-body physics arise, especially concerning the interplay between quantum interferences and the presence of interaction. In a quasi stationary context, mean-field studies \cite{Hartung2008PRL,PhysRevLett.100.033902,Hartmann2012AP} show that the presence of a nonlinearity in the wave equation describing the transport of ultracold bosonic atoms across a disordered region can give rise to an inversion of the coherent backscattering peak. On the other hand, many-body diagrammatic approaches \cite{Geiger2013} indicate that this inversion should be limited to a mean-field regime of very low atom-atom interaction strengths, while in a more realistic situation a dephasing is to be expected.

To shed more light on this issue, we propose to verify these observations in a most elementary setting which allows for numerical simulations beyond the mean-field Gross-Pitaevskii approach. Our system consists of two leads connected to a ring-shaped resonator that is threaded by a synthetic gauge flux. For such a system, it is well known that Aharonov-Bohm oscillations \cite{Ehrenberg1949,Aharonov1959,PhysRev.123.1511,WashburnWebb1986} take place in the transmission. The presence of disorder within the ring is further responsible for a cross-over from Aharonov-Bohm to Al'thsuler-Aronov-Spivak oscillations \cite{AAS1981,AASJETP1981,WashburnWebb1986,SharvinSharvin1981}, which is a mesoscopic phenomenon related to weak localisation. Al'tshuler-Aronov-Spivak oscillations have been investigated in some detail in mesoscopic physics, see e.g. Refs. \cite{MuratPRB1986,ImryPRL86,Doucot1987JP,PannetierPRB1985} and effects of electron-electron interaction were discussed \cite{AvishaiPRL96}.

In this paper, we shall investigate how Al'tshuler-Aronov-Spivak oscillations behave in the presence of bosonic interaction. To this end, we compute the disorder-averaged transmission of an interacting guided atom laser beam across a ring by means of a numerical integration of the Gross-Pitaevskii equation, and compare its findings with the predictions of a nonlinear diagrammatic theory. An inversion of the Al'tshuler-Aronov-Spivak oscillations profile in the disorder-averaged transmission is indeed encountered at small nonlinearities, in analogy with previous studies \cite{Hartung2008PRL,PhysRevLett.100.033902,Hartmann2012AP} on coherent backscattering. Finally, we use the truncated Wigner method which allows one to go beyond the mean-field approximation in order to verify to which extent this phenomenon prevails in the presence of finite atom-atom interaction strengths.

We start by presenting in Sec.~\ref{sec:sys} the guided atom laser configuration under study and the spatial discretisation scheme that we use to numerically solve the equations that describe our system. By representing the ring and the leads appearing in our system as a quantum graph, we are in a position to formulate in Sec.~\ref{sec:AASth} a theory explaining the appearance of Al'tshuler-Aronov-Spivak oscillations in the noninteracting case. We then present in Sec.~\ref{sec:nummeth} the numerical methods we use, namely the Gross-Pitaevskii equation and the truncated Wigner method. In Sec.~\ref{sec:res}, we shall first apply these methods to study Aharonov-Bohm oscillations in the presence of interaction. We then discuss the transition from Aharonov-Bohm to Al'tshuler-Aronov-Spivak oscillations and investigate how the latter are affected by the presence of interaction. Numerical findings on disorder-averaged transmission are compared with predictions of nonlinear diagrammatic theory which is described in Sec.~\ref{sec:diagth}, following the scheme developed in \cite{Wellens2006}.

\section{Description of the system \label{sec:sys}}
\begin{figure}
\begin{center}
\includegraphics[width=\linewidth, clip=true]{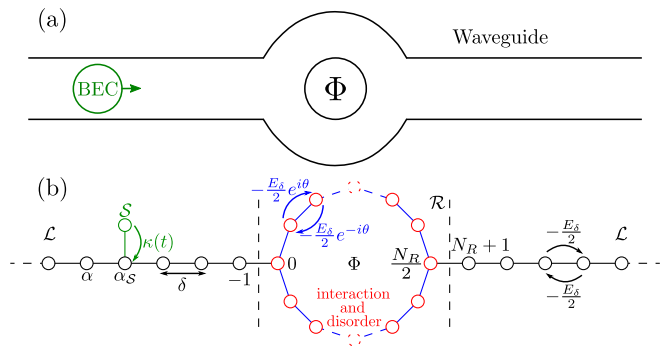}
\end{center}
\caption{(a) Sketch of the system under study. A Bose-Einstein condensate at temperature $T=0$ and chemical potential $\mu$ in a trap (sketched by a green circle) is outcoupled to a (semi-infinite) waveguide by for instance a radio-frequency knife or by multiphoton Raman transition. This waveguide is connected to another semi-infinite waveguide via a two-arms ring. In this ring is induced a tunable artificial gauge field $\Phi$. (b) Discretisation of the infinite 1D space, which is artificially subdivided into several regions labelled $\mathcal{L}$ (leads), $\mathcal{R}$ (ring) and $\mathcal{S}$ (source). Sites depicted in red exhibit both interaction and disorder. Edges depicted in blue exhibit, in addition to the Bose-Hubbard hopping term $-E_\delta/2$, a Peierls phase \cite{Peierls1933,PhysRevB.14.2239} shift $e^{\pm i\theta}$ depending on the direction of the rotation, due to an artificial gauge field. The phase shift $\theta$ acquired at each jump from one site to its neighbour is given by $\theta=\Phi/N_R$, with $N_R$ the number of sites within the ring.\label{fig:system}}
\end{figure}

The system we study is a Bose-Einstein condensate of $\mathcal{N}\rightarrow \infty$ particles at zero temperature $T=0$ and chemical potential $\mu$, which is outcoupled from a trap to a waveguide, e.g. by means of a radio-frequency knife \cite{Mewes1997PRL,Bloch1999PRL,Bolpasi2014NJP} or a multiphoton Raman transition \cite{Moy1997PRA,Hagley1999S,Robins2006PRL}, following the principle of an atom laser \cite{Bloch1999PRL,Guerin2006PRL,Couvert2008EEL,Gattobigio2009PRA,Gattobigio2011PRL,Vermersch2011PRA,Bolpasi2014NJP}. In this waveguide is engineered a two-arms ring, similar to an interferometer, in which is produced a synthetic gauge field \cite{Dalibard2011, Goldman2014} with tunable magnetic flux $\Phi$, the role of which is to break the symmetry between the two arms of the ring. Because of that flux, a different phase shift will be acquired by the atoms depending upon which arm is chosen to cross the ring. This symmetry breaking gives then rise to Aharonov-Bohm interference effects \cite{Ehrenberg1949,Aharonov1959,PhysRev.123.1511,WashburnWebb1986} scenario. Experimentally, such rings can be obtained by perpendicular intersection of red detuned lasers, as is explained in Refs. \cite{Amico2005, Ramanathan2011}. A horizontal atomic waveguide in a particular direction can be engineered by using a far-detuned laser beam, as in Ref. \cite{Guerin2006PRL}. The ring shaped geometry would then be connected to two semi-infinite leads, as is represented in \textsc{Fig.}\ \ref{fig:system}(a). 

The model to describe this transport process is provided by a system of evolution equations for the field operator $\hat{\phi}_S(t)$ of the source and the field operators $\hat{\psi}(x,t)$ of atoms within the waveguide structure, with $x$ representing positions in the leads and the ring. For a single infinite lead, we would have \cite{Ernst2010PRA}
\begin{align}
i\hbar \dfrac{\partial \hat{\psi}(x,t)}{\partial t} & = \hat{H}_0 \hat{\psi}(x,t)+ g(x) \hat{\psi}^\dagger(x,t)\hat{\psi}(x,t)\hat{\psi}(x,t) \nonumber \\
& \hspace{1cm} + K(x,t) \hat{\phi}_S \\
i\hbar \dfrac{\partial \hat{\phi}_S(t)}{\partial t} & = \mu \hat{\phi}_S + \int \mathrm{d}x K^*(x,t) \hat{\psi}(x,t),
\end{align}
with $K(x,t)$ the position-dependent coupling strength of the coupling between source and leads, $\mu$ the chemical potential of the source and $g(x)$ the effective one-dimensional interaction strength (which we assume to be present only inside the ring). The one-dimensional single-particle Hamiltonian (without the artificial gauge field) is given by 
\begin{equation}
\hat{H}_0 = \hat{H}_k + V(x) \quad \text{with} \quad \hat{H}_k = -\dfrac{\hbar^2}{2m} \dfrac{\partial^2}{\partial x^2}
\end{equation}
with $V(x)$ the disorder potential (which, like the interaction, is assumed to be present only inside the ring).

In view of implementing the truncated Wigner method, we discretise the one-dimensional space, as shown in \textsc{Fig.}~\ref{fig:system}(b), in a series of sites labelled by an index $\alpha$ and spaced by $\delta$. We describe the kinetic energy operator in terms of a finite-difference scheme
\begin{equation*}
\dfrac{\partial^2}{\partial x^2} \psi(x) \simeq \dfrac{\psi(x+\delta) + \psi(x-\delta) - 2 \psi(x)}{\delta^2}.
\end{equation*} 
Through the discretisation, each site obtains an on-site energy $E_\delta = \hbar^2/m\delta^2$, with $\hbar$ the reduced Planck constant and $m$ the mass of the atoms, and a nearest-neighbour hopping $E_\delta/2$.

We define several space regions in our discretisation, namely $\mathcal{L}$ for the leads, $\mathcal{R}$ for the ring and $\mathcal{S}$ for the source. Smooth exterior complex scaling \cite{Balslev1971CMP, Simon1979PLA, Simon1973AM, Junker1982AAMP, Reinhardt1982ARPC, Ho1983PR, Loewdin1988AQC, Moiseyev1998PR} is used according to Ref.\ \cite{Dujardin2015PRA} for absorption at both ends of the leads to avoid artifacts due to the finite extension of the leads in the numerical treatment. The effective Bose-Hubbard Hamiltonian describing this system can be decomposed in four sub-Hamiltonians
\begin{equation}
\hat{H} = \hat{H}_\mathcal{L} + \hat{H}_\mathcal{LR} + \hat{H}_\mathcal{R} + \hat{H}_\mathcal{S} 
\vspace{-0.25cm}
\end{equation}
with
\begin{align}
& \hat{H}_\mathcal{L} = \sum_{\alpha \in \mathcal{L}} \Big[ E_\delta \hat{a}_\alpha^\dagger \hat{a}_\alpha - \dfrac{E_\delta}{2} \left( \hat{a}_{\alpha+1}^\dagger \hat{a}_\alpha + \hat{a}_\alpha^\dagger\hat{a}_{\alpha+1} \right) \Big] \nonumber \\ 
& \hat{H}_\mathcal{LR} = -\dfrac{E_\delta}{2} \left( \hat{a}_{-1}^\dagger \hat{a}_0 + \hat{a}_{0}^\dagger \hat{a}_{-1} \right. \nonumber \\
&  \hspace{2.5cm} \left.+ \hat{a}_{N_R/2}^\dagger \hat{a}_{N_R+1} + \hat{a}_{N_R+1}^\dagger \hat{a}_{N_R/2} \right) \nonumber \\ 
& \hat{H}_\mathcal{R} = \sum_{\alpha \in \mathcal{R}} \Big[\left(E_\delta + V_\alpha \right) \hat{a}_\alpha^\dagger \hat{a}_\alpha \nonumber \\
& \hspace{1cm}- \dfrac{E_\delta}{2} \left( \hat{a}_{\alpha-1}^\dagger \hat{a}_\alpha e^{i\theta} + \hat{a}_{\alpha+1}^\dagger \hat{a}_\alpha e^{-i\theta} \right) + \dfrac{g}{2} \hat{a}_\alpha^\dagger \hat{a}_\alpha^\dagger \hat{a}_\alpha \hat{a}_\alpha \Big] \nonumber \\
& \hat{H}_\mathcal{S} = \kappa(t)\hat{a}^\dagger_{\alpha_\mathcal{S}}\hat{b} + \kappa^*(t)\hat{b}^\dagger\hat{a}_{\alpha_\mathcal{S}} + \mu \hat{b}^\dagger \hat{b},
\label{eq:Hamiltonian}
\end{align}
each of which being associated with the corresponding region of the space they are labelled by.

In this Hamiltonian, we have introduced by $\hat{a}_\alpha^\dagger$ and $\hat{a}_\alpha$ the creation and annihilation operators at site $\alpha$ and by $\hat{b}^\dagger$ and $\hat{b}$ the creation and annihilation operators of the source which is maintained at chemical potential $\mu$ and $T=0$ \footnote{Note that within the Hamiltonian $\hat{H}_{\mathcal{R}}$ we have identified the sites $\alpha_{N_R+1}\equiv\alpha_0$ and $\alpha_{-1} \equiv \alpha_{N_R}$ to simplify notation. They should not be confused with sites in the leads carrying the same labels.}. We treat this source as a Bose-Einstein condensate containing $\mathcal{N} \rightarrow \infty$ atoms and make the approximation that it is connected to one single lattice site labelled by $\alpha_{\mathcal{S}}$. The coupling $\kappa(t)$ between the source and the leads is smoothly ramped on with time (for instance by varying the intensity of the radio-frequency field in case of a radio-frequency knife) and approaches a maximal value. This latter tends to zero such that $\mathcal{N}\vert \kappa(t)\vert^2$ remains constant \cite{Guerin2006PRL,Riou2008PRA,Dujardin2014APB}, which implies that the number of atoms in the scattering region remains constant, too. In this limit, a stationary many-body scattering state can therefore be achieved. In the case where the source is connected, as described, to an infinite lead, which amounts to considering $\hat{H} = \hat{H}_\mathcal{S} +\hat{H}_\mathcal{L}$, then it would inject a free flux of atoms yielding a stationary density and current given by \cite{Dujardin2015PRA,Dujardin2014APB}
\begin{align}
\rho^\varnothing & = \dfrac{1}{\delta} \dfrac{\mathcal{N}\vert \kappa(t)\vert^2}{\mu(2E_\delta - \mu)} \\
j^\varnothing & = \dfrac{1}{\hbar} \dfrac{\mathcal{N}\vert \kappa(t)\vert^2}{\sqrt{\mu(2E_\delta - \mu)}}. 
\label{eq:denscurr}
\end{align}

The on-site interaction strength is controlled by the parameter \cite{PhysRevLett.81.938} $g = 2\hbar \omega_\perp a_S/\delta$, $\omega_\perp$ being the perpendicular confinement frequency of the trap and $a_S$ the s-wave scattering length. Finally, disorder is brought into the system through the on-site parameters $V_\alpha$. The disorder we use is in the continuous space generated by \cite{Paul2009AndLocPRA,Dujardin2016}
\begin{equation}
V(x) = \bar{V}_0 \int \dfrac{1}{\sqrt{\sigma\sqrt{\pi}}} \exp\left[-\frac{(x-y)^2}{2\sigma^2}\right] \eta(y) \mathrm{d} y,
\end{equation}
where $\bar{V}_0$ is the amplitude of the disorder and $\sigma$ its correlation length. The correlator $\eta(y)$ is a gaussian random white noise with zero mean and unit variance, \emph{i.e.}\ $\langle \eta(x)\eta(y) \rangle = \delta(x-y)$, with $\langle \cdot \rangle$ the random average. In the framework of the above discretisation scheme, disorder is then represented by the on-site energies
\begin{equation}
V_\alpha = \bar{V}_0 \sum_{\alpha' = 0}^{N_R} \dfrac{1}{\sqrt{\sigma\sqrt{\pi}}} \exp\left[-\frac{\delta^2}{2\sigma^2}(\alpha-\alpha')^2\right] \eta_{\alpha'}.
\label{eq:disorder}
\end{equation}
This imposes a condition on the discretisation, namely $\delta \ll \sigma$, in order that the discretisation scheme captures the details of the disorder. 

\section{Theory of Al'tshuler-Aronov-Spivak oscillations \label{sec:AASth}}
Let us first consider the noninteracting case, that is a ring penetrated by an Aharonov-Bohm artificial gauge flux in the presence of disorder but without any interaction. The ring and the leads can be represented as a quantum graph \cite{PhysRevLett.79.4794,KOTTOS199976,PhysRevLett.85.968} with two vertices, two internal bonds of finite length and two external bonds of infinite extension. On this graph, the Green function can be represented as a sum over all possible paths $\gamma$ linking two given points $\alpha',\alpha$ on the graph (see Appendix \ref{sec:GFprod}),
\begin{equation}
G(\alpha,\alpha',\mu) = \dfrac{1}{i E_\delta \sin(k \delta)} \sum_\gamma A_\gamma e^{iS_\gamma/\hbar},
\label{eq/GF_SumSCtraj}
\end{equation}
with $k\delta = \arccos\left(1 - \mu/E_\delta\right) \simeq \sqrt{2 \mu/E_\delta}$ for $0 < \mu/E_\delta \ll 1$. In the above equation, $S_\gamma$ is the accumulated action integral along the path $\gamma$ and the prefactor $A_\gamma = r^{n_r}t^{n_t}$ is the product of reflection and transmission matrix elements at each junction that a trajectory encounters, where $n_r$ (resp. $n_t$) is the number of reflections (resp. transmissions) along the path $\gamma$. Those matrix elements can be obtained from the analysis of a scattering problem across a symmetric Y junction, in the absence of disorder. Denoting by $E_Y$ the energy on the junction sites, which may be assumed to be different from the other on-site energies $E_\delta$, we obtain 
\begin{align}
r = - \dfrac{1 - E_Y/E_\delta + \frac{1}{2} e^{ik\delta}}{1 - E_Y/E_\delta + e^{ik\delta}- \frac{1}{2} e^{-ik\delta}}, \\
t = \dfrac{i\sin k\delta}{1 - E_Y/E_\delta + e^{ik\delta}- \frac{1}{2} e^{-ik\delta}}.
\end{align}
Those probability amplitudes satisfy the continuity and conservation of current equations
\begin{align}
1+r & = t \\
\vert r\vert^2 + 2 \vert t \vert^2 & = 1.
\end{align}
In the continuous limit where the spacing $\delta \rightarrow 0$ vanishes, a nonvanishing transmission is obtained only for $E_Y = \frac{3}{2}E_\delta$, which yields
\begin{align}
r & = - \dfrac{e^{ik\delta}-1}{2e^{ik\delta} - 1 - e^{-ik\delta}} \overset{k\delta\rightarrow 0}{\longrightarrow} -\dfrac{1}{3}, \label{eq:rcl}  \\
t & = \dfrac{e^{ik\delta} - e^{-ik\delta}}{2e^{ik\delta} - 1 - e^{-ik\delta}} \overset{k\delta\rightarrow 0}{\longrightarrow} \dfrac{2}{3}. \label{eq:tcl}
\end{align}
Here, we choose $E_Y = E_\delta$, which corresponds to the case of a nearly closed ring (actually with nearly disconnected arms) that is weakly connected to the waveguides, as well as $\mu = 0.2 E_\delta$. This yields the following expressions for $r$ and $t$:
\begin{align}
r & = - \dfrac{e^{ik\delta}}{2e^{ik\delta} - e^{-ik\delta}} = -\dfrac{43}{97}+\dfrac{24i}{97} \label{eq:rwc}  \\
t & = \dfrac{e^{ik\delta} - e^{-ik\delta}}{2e^{ik\delta} - e^{-ik\delta}} = \dfrac{54}{97}+\dfrac{24i}{97} \label{eq:twc}.
\end{align}
The disorder within the ring is supposed to be weak and smooth, with an amplitude $\bar{V}_0 \ll \mu$ and a spatial correlation length $\sigma$ satisfying $k\sigma \gg 1$ as well as $\sigma \gg \delta$ such that reflections inside each arm of the ring can safely be neglected. The action integral within each arm can then be written as
\begin{align}
S & = \hbar \sum_{\alpha=1}^{N_R} \arccos \left( 1 - \dfrac{\mu - V_\alpha}{E_\delta} \right) \simeq \hbar \int_0^{N_R} \sqrt{\dfrac{\mu-V_\alpha}{2E_\delta}} \mathrm{d}\alpha \\
& \simeq N_R \hbar \sqrt{\dfrac{\mu}{E_\delta}} - \dfrac{\hbar}{2\sqrt{\mu E_\delta}} \int_0^{N_R} V_\alpha \mathrm{d}\alpha,
\end{align}
which means that the phase factors for the upper branch $\exp(iS_u/\hbar)$ and the lower branch $\exp(iS_d/\hbar)$ can safely be considered as independent complex random numbers with unit norm, provided that the ring is sufficiently long such that 
\begin{equation}
\overline{\left\vert\int_0^{N_R} V_\alpha \mathrm{d}\alpha\right\vert} \gg \pi \sqrt{\mu E_\delta}.
\end{equation}
Now if we consider the presence of an Aharonov-Bohm flux $\Phi$ within the ring, we can write
\begin{equation}
\exp\left(\frac{i}{\hbar}S_u\right) = \exp\left(i\left(\Phi_u \pm \frac{\Phi}{2}\right)\right) 
\end{equation}
and
\begin{equation}
\exp\left(\frac{i}{\hbar}S_d\right) = \exp\left(i\left(\Phi_d \mp \frac{\Phi}{2}\right)\right),
\end{equation}
for the upper and lower arm, respectively, where the upper (resp. lower) sign is associated with the path from the left to the right (resp. the right to the left) junction and $\Phi_u,\Phi_d$ are random phases accounting for disorder in upper or lower arms of the ring. 

Reflection and transmission amplitudes across the ring can then be obtained from the Green function $G(\alpha,\alpha_\mathcal{S},\mu)$, with $\alpha$ being located in the lead before or behind the ring for reflection and transmission, respectively. More precisely, these amplitudes are obtained by normalizing these Green functions with respect to the Green function of free motion along a clean 1D lattice, namely
\begin{equation}
G_0 = \dfrac{e^{ik(\alpha-\alpha_\mathcal{S})}}{iE_\delta \sin(k\delta)}.
\end{equation}
The probability amplitude of reflection $R$ can therefore be written as a sum of probability amplitudes associated with trajectories yielding a reflection after a possible complicated journey in the ring. At each Y junction, such an amplitude is multiplied either by $r$ in case of reflection or by $t$ in case of transmission across this junction. In addition, it is also multiplied by the phase factor accounting for its journey within the ring. Each exploration of the upper (resp. lower) branch yields a $\exp(i\Phi_u)$ (resp. $\exp(i\Phi_d)$) phase factor and each trip in the clockwise (resp. counterclockwise) direction yields a $e^{i\Phi}$ (resp. $e^{-i\Phi}$) phase factor.

\begin{figure}
\begin{center}
\includegraphics[width=8cm, clip=true]{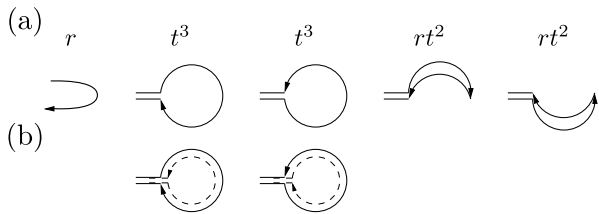}
\end{center}
\caption{(a) Schematic representation of the diagrams yielding an expression for the reflection up to corrections of power 5 in $r$ and $t$, as is shown in Eq.~\eqref{eq:refl}. Since at each junction either a reflection or a transmission event occurs, we have to multiply each trajectory amplitude either by $r$ or $t$, depending on which event took place. An additional phase factor $\exp\left[i(\Phi_{u/d} \pm \Phi)\right]$ should also be considered as a result of the exploration of the upper or lower branch in the clockwise or counterclockwise direction of rotation. (b) Sketch of the intensity diagrams that give rise to the leading-order terms beyond the diagonal approximation. The two paths corresponding to the wavefunction (solid line) and its complex conjugate (dashed line) interfere constructively (destructively) for fluxes $\Phi$ that are an even (odd) multiple of $\pi/2$.}
\label{fig:foR}
\end{figure}

Up to some (unimportant) global phase factor, we have for the reflection amplitude
\begin{align}
R & = r + t^2r\left(e^{2i\Phi_u}+e^{2i\Phi_d}\right) \nonumber \\
& \hspace{0.65cm} + t^3 e^{i(\Phi_u + \Phi_d)}\left(e^{i\Phi} + e^{-i\Phi}\right) + \mathcal{O}((r,t)^5)
\label{eq:refl}
\end{align}
which is graphically illustrated in \textsc{Fig.} \ref{fig:foR} (a). 
For the transmission, we find
\begin{align}
T & = t^2 \left(e^{i\left(\Phi_u + \Phi/2\right)} + e^{i\left(\Phi_d -\Phi/2\right)}\right) \nonumber \\
& \hspace{0.65cm} + t^2r^2 \left(e^{3i\Phi_u} + e^{3i\Phi_d}\right) \nonumber \\
& \hspace{0.65cm} + 2t^3r \left(e^{2i\Phi_u}e^{i\left(\Phi_d - \Phi/2\right)} + e^{2i\Phi_d}e^{i\left(\Phi_u + \Phi/2\right)}\right) \nonumber \\ 
& \hspace{0.65cm} + t^4 \left(e^{2i\Phi_u}e^{i\left(\Phi_d + 3\Phi/2\right)}\right) +\left( e^{2i\Phi_d}e^{i\left(\Phi_u - 3\Phi/2\right)}\right) \nonumber \\
& \hspace{0.65cm} + \mathcal{O}((r,t)^6),
\end{align}
up to contributions of higher order in the powers of $r$ and $t$. Reflection and transmission probabilities are obtained from the amplitudes $R$ and $T$ by calculating their modulus square. They thus involve double sums over trajectories. To observe Al'tshuler-Aronov-Spivak oscillations in these probabilities, a further average over disorder is required. From the original double sum over trajectories only those pairs of trajectories survive this disorder average that have zero net power of the complex random numbers $e^{i\Phi_u}$ and $e^{i\Phi_d}$, as can be seen in Fig. \ref{fig:foR}(b). The disorder-averaged reflection and transmission then write
\begin{align}
\overline{\vert R \vert^2} & = \vert r \vert^2 + 2 \vert t \vert^4 \vert r \vert^2 + 2 \vert t \vert^6 (1 + \cos 2 \Phi) + \mathcal{O} ((r,t)^{10}) \\
\overline{\vert T \vert^2} & = 2 \vert t \vert^4 + 2 \vert t \vert^4 \vert r \vert^4 + 8 \vert t \vert^6 \vert r \vert^2 + 2 \vert t \vert^8 \nonumber \\
& \hspace{0.65cm} + 4 \vert t \vert^6 (tr^* + rt^*) \cos 2 \Phi + \mathcal{O}((r,t)^{12}) \label{eq:trans}.
\end{align}
In those expressions, pairs of paths that could give rise to a $\cos(\Phi)$ contribution display a non zero net power of the complex random numbers $e^{i\Phi_u}$ and $e^{i\Phi_d}$. They do not survive the disorder averaging. Evaluating in the expression \eqref{eq:trans} for the transmission
\begin{equation}
tr^* + rt^* = - \dfrac{\sin^2 k\delta}{\vert 1 - \frac{E_Y}{E_\delta} + e^{ik\delta} - \frac{1}{2} e^{-ik\delta} \vert^2},
\end{equation}
we see that $\cos(2\Phi)$ oscillations of reflection probability are compensated on transmitted side with nearly the same magnitude (at this level of approximation). We specifically obtain for $E_Y = E_\delta$
\begin{equation}
2 (tr^* + rt^*) = -\dfrac{72}{97} \simeq -0.8.
\end{equation}
This robust enhancement of the reflection, along with the associated drop in the transmission, arises due to coherent backscattering and is a clear signature of weak localisation. Similar results have been obtained for quasi one-dimensional disordered electronic systems in the presence of a magnetic field \cite{Doucot1987JP} or in two-dimensional arrays of nonsuperconducting metallic rings \cite{PannetierPRB1985}. While such oscillations are usually encountered in quasi one-dimensional models \cite{AAS1981,AASJETP1981,WashburnWebb1986,SharvinSharvin1981,MuratPRB1986,ImryPRL86,Doucot1987JP,PannetierPRB1985}, we chose here to restrict the motion to exactly one dimension (i.e. we consider a waveguide that does not allow for the population of excited transverse modes at the chemical potential under consideration) since the use of ultracold gases make this choice possible. 

\section{Numerical methods \label{sec:nummeth}}

\subsection{Mean-field Gross-Pitaevskii approach}
The Gross-Pitaevskii approximation has been used for the numerical simulation of atom-laser scenarios \cite{Leboeuf2001PRA,Carusotto2001PRA,Paul2005PRL,Paul2005PRA,Paul2007PRA} and it was proven \cite{PhysRevA.61.043602} that it is a good approximation in the limit of a large atomic density and small interaction strength. Starting from our Hamiltonian given in Eq.~\eqref{eq:Hamiltonian} and working in the Heisenberg picture, we obtain the evolution of the annihilation operators according to
\begin{align}
i\hbar \dfrac{\partial \hat{a}_\alpha(t)}{\partial t} & = (E_\alpha + V_\alpha) \hat{a}_\alpha(t) - \sum_{\alpha'} J_{\alpha\alpha'} \hat{a}_{\alpha'}(t)\nonumber\\
& \hspace{0.5cm} + g_\alpha \hat{a}_\alpha^\dagger(t) \hat{a}_\alpha(t) \psi_\alpha(t) + \kappa(t) \delta_{\alpha,\alpha_{\mathcal{S}}} \hat{b}(t) \\
i\hbar \dfrac{\partial \hat{b}(t)}{\partial t} & = \mu \hat{b}(t) + \kappa^*(t)\hat{a}_{\alpha_{\mathcal{S}}}(t),
\end{align}
where $J_{\alpha\alpha'}$ encodes the matrix elements describing hopping from one site to another within in the leads, the ring and the junction. Additionally, they include the Peierls phase within the ring. 

The mean-field limit consists in the regime where the on-site densities are large and the interaction strength is weak. This allows one to replace the quantum operators by c-numbers. In that limit, the dynamics of the system is governed by the Gross-Pitaevskii equation 
\begin{align}
i\hbar \dfrac{\partial \psi_\alpha(t)}{\partial t} & = (E_\alpha + V_\alpha-\mu) \psi_\alpha(t) - \sum_{\alpha'} J_{\alpha\alpha'} \psi_{\alpha'}(t) \nonumber\\
& \hspace{0.5cm} + g_\alpha \vert\psi_\alpha(t)\vert^2 \psi_\alpha(t) + \kappa(t) \delta_{\alpha,\alpha_{\mathcal{S}}} \chi(t) \\
i\hbar \dfrac{\partial \chi(t)}{\partial t} & = \kappa^*(t)\psi_{\alpha_{\mathcal{S}}}(t),
\label{eq:GP}
\end{align}
where we have made the ansatz $\psi_\alpha(t) = \langle \hat{a}_\alpha \rangle e^{-i\mu t}$ and $\chi(t) = \langle \hat{b} \rangle e^{-i\mu t}$ with $\psi_\alpha(t_0) = 0$ and $\chi(t_0) = \sqrt{\mathcal{N}}$, corresponding to empty waveguides, an empty ring, and a coherent Bose-Einstein condensate within the reservoir of atoms. 

It is clear from Eq.~\eqref{eq:GP} that $\chi(t) = \sqrt{\mathcal{N}} (1 + \mathcal{O}(\vert\kappa\vert^2))$ for some finite time $t-t_0$. Therefore, in the limit where the coupling $\kappa(t)$ tends to zero in such a manner that $\mathcal{N} \vert\kappa(t)\vert^2$ remains constant, we can neglect the time evolution of $\chi(t)$ and we are left with a nonlinear Schrödinger equation containing an additional source term \cite{Paul2005PRL,Paul2007PRA,Ernst2010PRA}
\begin{align}
i\hbar \dfrac{\partial \psi_\alpha(t)}{\partial t} & = (E_\alpha + V_\alpha-\mu) \psi_\alpha(t) - \sum_{\alpha'} J_{\alpha\alpha'} \psi_{\alpha'}(t)\nonumber\\
& \hspace{0.5cm} + g_\alpha \vert\psi_\alpha\vert^2 \psi_\alpha(t) + \kappa(t) \delta_{\alpha,\alpha_{\mathcal{S}}} \sqrt{\mathcal{N}},
\label{eq:SchrSourterm}
\end{align}
The on-site density and current are defined as
\begin{align}
n_\alpha & = |\psi_\alpha|^2, \\
j_\alpha & = \dfrac{iE_\delta}{2\hbar}\left[\psi_{\alpha+1}^*(t)\psi_\alpha(t) - \psi_{\alpha}^*(t)\psi_{\alpha+1}(t)\right].
\end{align}
The main drawback of this approach is that, notably when disordered potentials are considered, even a weak atom-atom interaction can generate the population of a non-condensed cloud off the energy shell through two-body scattering \cite{Dujardin2016,Geiger2012PRL,Geiger2013NJoP}. Those effects are beyond the scope of the mean-field Gross-Pitaevskii approach and must be adressed by means of another method.

\subsection{Truncated Wigner method}
The truncated Wigner method \cite{Wigner1931,Wigner1932PR,Moyal1949PCPS,Steel1998PRA,Sinatra2002JPBAMOP,Polkovnikov2003PRA}, which has been successfully adapted to the context of an atom-laser scenario \cite{Dujardin2015PRA,DujardinAdP2015}, allows one to go beyond the mean-field approximation described by the Gross-Pitaevskii approach. The principle of the method consists in sampling the many-body quantum state of the system by classical fields $\displaystyle \{\psi_\alpha\}_{\alpha\in\mathcal{R},\mathcal{L}}$ that properly represent the initial state of the system at the initial time $t_0$ and evolve according to a slightly modified Gross-Pitaevskii equation. As we consider that initially, at $t=t_0$, the waveguides and the ring are empty whilst the reservoir is populated with a large number $\mathcal{N}$ of atoms, we can decouple the initial Wigner function of the system as
\begin{align}
\mathcal{W}(\{\psi_\alpha,\psi_\alpha^*\},t_0) & = \mathcal{W}_G(\{\psi_\alpha,\psi_\alpha^*\},t_0) \nonumber \\
& \hspace{0.5cm}\times \mathcal{W}_S(\chi,\chi^*,t_0),
\end{align}  
that is, as a product of the Wigner functions describing the source and the scattering system. Initially, the waveguides and the ring are empty, which implies that their Wigner function can be written as a product of vacuum states
\begin{equation}
\mathcal{W}_G(\{\psi_\alpha,\psi_\alpha^*\},t_0) = \prod_{\alpha} \left(\frac{2}{\pi}\right) e^{-2|\psi_\alpha|^2}.
\end{equation}
In practice, the classical field amplitudes are determined as
\begin{equation}
\displaystyle \psi_\alpha(t=t_0) = \dfrac{1}{2}\left(\mathcal{A}_\alpha + i \mathcal{B}_\alpha\right)
\end{equation}
where $\mathcal{A}_\alpha$ and $\mathcal{B}_\alpha$ are real and independent gaussian random variables fulfilling
\begin{align}
\overline{\mathcal{A}_\alpha}  &=  \overline{\mathcal{B}_\alpha} = 0, \\
\overline{\mathcal{A}_{\alpha'}\mathcal{A}_\alpha}  &= \overline{\mathcal{B}_{\alpha'}\mathcal{B}_\alpha}  = \delta_{\alpha',\alpha}, \\
\overline{\mathcal{A}_{\alpha'}\mathcal{B}_\alpha} &= 0,
\end{align}
in which the notation $\overline{\cdot}$ means that an average over the random variables is performed. Because of that, each site of the system but the source exhibits an artificial nonzero average population $\overline{\vert\psi_\alpha(t_0)\vert^2} = 1/2$ which one has to subtract when computing the atomic density. 

As the source is assumed to be populated with a large number $\vert\chi\vert^2=\mathcal{N} \gg 1$ of atoms, the Wigner function of the source can be considered as that of a coherent state
\begin{equation}
\mathcal{W}_\mathcal{S}(\chi,\chi^*,t_0) = \left(\dfrac{2}{\pi}\right) e^{-2|\chi - \sqrt{\mathcal{N}}|^2}.
\end{equation}
This very high number of atoms is such that the relative uncertainties on the amplitude and the phase of the source are negligible; we can then treat the source in a classical manner so that $\displaystyle \chi(t=t_0) = \sqrt{\mathcal{N}}$. If, in addition, the coupling $\kappa(t)$ is chosen such that $\kappa(t)\rightarrow 0$ in such a manner that $\mathcal{N}\vert\kappa\vert^2$ remains finite, the depletion of the source or any back-action of the waveguide on the source can be safely neglected \cite{Dujardin2015PRA} and one can solely focus on the evolution within the waveguides and the ring. In this case, the propagation equation for the amplitude on each sampling point is given by
\begin{align}
i\hbar \dfrac{\partial \psi_\alpha}{\partial t} & = (E_\delta - \mu + V_\alpha) \psi_\alpha + \sum_{\alpha'} J_{\alpha\alpha'} \psi_{\alpha'} \nonumber\\
& \hspace{1cm} + g_\alpha \left( \vert\psi_\alpha\vert^2 -1 \right) \psi_\alpha + \kappa(t) \sqrt{\mathcal{N}} \delta_{\alpha,\alpha_\mathcal{S}},
\end{align}
where $J_{\alpha\alpha'}$ are the hopping matrix elements from site $\alpha$ to site $\alpha'$. 

Observables are computed through an average over the random initial conditions. This, for instance, yields for the on-site density and current
\begin{align}
n_\alpha & = \overline{|\psi_\alpha|^2} - \frac{1}{2}, \\
j_\alpha & = \dfrac{iE_\delta}{2\hbar}\overline{\psi_{\alpha+1}^*(t)\psi_\alpha(t) - \psi_{\alpha}^*(t)\psi_{\alpha+1}(t)},
\end{align}
where the subtraction of $1/2$ in the density compensates for the artificial $1/2$ atom per site, as explained above. 

The truncated Wigner method allows one, in great contrast to a mean-field approach, to access both coherent and incoherent quantities. The coherent contributions to the on-site density and the current are given by
\begin{align}
n_\alpha^{\text{coh}} & = \left|\overline{\psi_\alpha}\right|^2, \\
j_\alpha^{\text{coh}} & = \dfrac{iE_\delta}{2\hbar}\left(\overline{\psi_{\alpha+1}^*(t)} \, \overline{\psi_\alpha(t)} - \overline{\psi_{\alpha}^*(t)} \, \overline{\psi_{\alpha+1}(t)}\right),
\end{align}
and the incoherent ones are then obtained through
\begin{align}
n_\alpha^{\text{incoh}} & = n_\alpha - n_\alpha^{\text{coh}}, \\
j_\alpha^{\text{incoh}} & = j_\alpha - j_\alpha^{\text{coh}}.
\end{align}
The transmission is defined as the ratio between the current at a given site in the downstream region behind the ring and the free stationary current
\begin{equation}
\vert T \vert^2 = \lim_{t \rightarrow \infty} j(t)/j^\varnothing.
\end{equation}
The reflection is obtained by considering that
\begin{equation}
\vert R \vert^2 + \vert T \vert^2 = 1. 
\end{equation}
Similarly as for the density and the current, one can also introduce the coherent and incoherent part of the transmission as
\begin{align}
T^{\text{coh}} & = \lim_{t \rightarrow \infty} j(t)^{\text{coh}}/j^\varnothing, \\
T^{\text{incoh}} & = \lim_{t \rightarrow \infty} j(t)^{\text{incoh}}/j^\varnothing.
\end{align}

\section{Results \label{sec:res}}

\subsection{Aharonov-Bohm oscillations}
We begin our numerical study by setting the disorder strength to zero, which allows us to focus solely on the interplay of interference and interaction. 
This is the standard Aharonov-Bohm scenario where the interference between the semiclassical contributions resulting for the two arms can be constructive or destructive depending on the artificial flux $\Phi$ within the ring. This can be verified by computing the transmission, which is defined by the ratio between the current at a site located after the ring and the injected free current $j^\varnothing$ defined in Eq.~\eqref{eq:denscurr}.

\begin{figure}[h!]
\includegraphics[width=\linewidth]{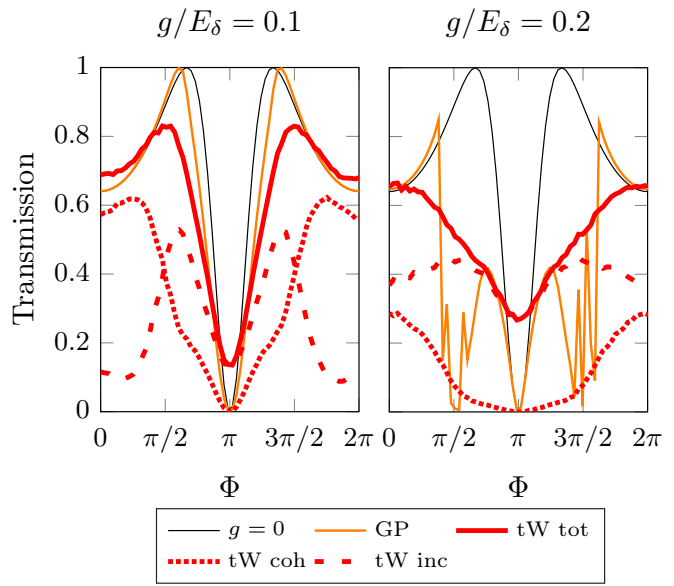}
\caption{Noninteracting ($g/E_\delta = 0$, thin black curve), Gross-Pitaevskii (orange/light grey curve) and truncated Wigner (red/heavy grey curves) simulations showing the transmission versus $\Phi$ with $\mu/E_\delta = 1$, $g/E_\delta=0.1$ (left panel) and  $g/E_\delta=0.2$ (right panel), $\sqrt{\mathcal{N}} \vert \kappa \vert/E_\delta = 1$. A tiny ring with $N_R=6$ sites was considered. The noninteracting (black) curve shows the typical Aharonov-Bohm oscillations as a result of interferences at the exit of the ring. Perfect transmission is reached for $\Phi\simeq 3\pi/5$ and a transmission blockade is observed at $\Phi = \pi$. While Gross-Pitaevskii simulation essentially confirm this behaviour for weak interaction (left panel), the truncated Wigner curves show that perfect transmission is inhibited and the transmission blockade at $\Phi = \pi$ is removed, as a result of creation of incoherent particles within the ring. For stronger interactions (right panel), we also observe oscillations in the Gross-Pitaevskii transmission (orange/light grey curve) due to bistability that indicates a breakdown of matter wave coherence. This is confirmed by truncated Wigner simulations yielding dominantly incoherent contributions.}
\label{fig:AB} 
\end{figure}
The expected interference pattern is confirmed in \textsc{Fig.}~\ref{fig:AB}. The noninteracting (black) curve shows the steady state result of a simulation of the transmission as a function of the artificial flux $\Phi$ for $g=0$. For $\Phi\simeq 3\pi/5$, the transmission of atoms is perfect and reaches one, which is a signature of destructive interference of the reflection at the entry side. On the other hand, the transmission blockade at $\Phi = \pi$ highlights destructive interferences at the exit site of the ring, namely between the partial waves crossing each arm of the ring, giving rise to a transmission blockade. Performing a Gross-Pitaevskii simulation for small interaction ($g/E_\delta=0.1$, orange/light grey curve), we observe a displacement of the maxima but no lifting of the transmission blockade. Truncated Wigner simulations reveal that a suspension of the blockade occurs at $\Phi = \pi$. This is entirely due to the incoherent part of the transmission, which could not have been predicted by Gross-Pitaevskii simulations. Interaction is responsible for this incoherent transmission because non condensed particles, with kinetic energy slightly lower or higher compared to that of the condensate, are created within the ring as a result of interaction. 

The right panel of \textsc{Fig.}~\ref{fig:AB} shows the transmission for larger $g$. Oscillations in the Gross-Pitaevskii curve can be seen as an artifact of the mean-field approach. Those oscillations are a signature of bistability, as documented and observed in \cite{Paul2005PRL,Paul2007PRA,Dujardin2015PRA}. Truncated Wigner simulations reveal a breakdown of matter wave coherence. Aharonov-Bohm-like oscillations are nevertheless encountered owing to significant remnants of coherent components of the atomic cloud near $\Phi = 0$ and $\Phi = 2\pi$. For stronger interactions, Aharonov-Bohm oscillations are expected to be washed out according to the study undertaken in Ref.\ \cite{Haug2017}.

\subsection{From Aharonov-Bohm to Al'tshuler-Aronov-Spivak oscillations}
As explained in Sec.~\ref{sec:AASth}, if we add a smooth (for instance gaussian-correlated) disorder potential within the ring, we can cancel the Aharonov-Bohm oscillations in the transmission and reveal Al'tshuler-Aranov-Spivak ones \cite{AAS1981,AASJETP1981,WashburnWebb1986,SharvinSharvin1981}. Indeed, a random phase is acquired after a trip in the ring due to the fact that the disorder potentials in the upper and lower arm of the ring are not correlated with each other. For pairings of trajectories that provide contributions to Aharonov-Bohm oscillations, the phase averages out and such pairings do not contribute on average. On the other hand, as pairings of trajectories that provide contributions to Al'tshuler-Aronov-Spivak oscillations, which are time-reversed conjugates of each other, are such that the same phase is accumulated; such pairings do not cancel with each other and are preserved after averaging. 

In order to develop a nonlinear diagrammatic theory taking into account interaction effects, we want to work in the semiclassical regime which corresponds to an action $S \gg \hbar$ and a correlation length $\sigma \gg \lambda$. For that purpose, we need to simultaneously enforce the four conditions 
\begin{equation}
\delta \ll \lambda \ll \sigma \ll L \ll l_{\text{loc}},
\end{equation}
where $l_{\text{loc}} \propto \exp(4k^2\sigma^2)$ \cite{Paul2009AndLocPRA} is the localisation length for strong (Anderson) localisation \cite{Anderson1958PR} within an arm of the ring, and $L$ the length of the ring. Furthermore, $\mu$ has to be small compared to $E_\delta$ to be close to the free dispersion relation of the continuous one-dimensional space. Specifically, we choose the chemical potential $\mu/E_\delta = 0.2$, the disorder amplitude $\bar{V}_0=0.0238$ and the correlation length $\sigma = 20 \delta$. We indeed have $k\delta \approx 0.67 < 1$ and $k\sigma \approx 13.4 \gg 1$, indicating that we are working in the validity regime of semiclassical methods. 

\begin{figure}[h!]
\includegraphics[width=1.05\linewidth, clip=true]{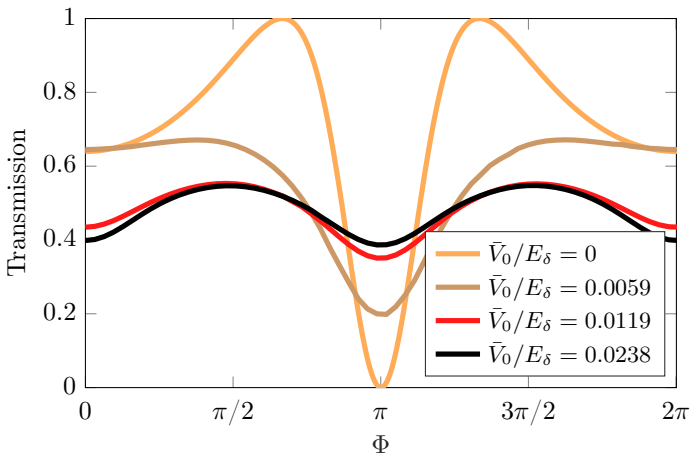}
\caption{Noninteracting simulation showing the transmission versus $\Phi$ with $\mu/E_\delta = 0.2$ and $N_R = 200$ sites. We considered 20000 realisations of a gaussian correlated disorder according to Eq. \eqref{eq:disorder} with the amplitude $\bar{V}_0$ and correlation length $\sigma = 20 \delta$. Aharonov-Bohm oscillations (orange/light grey curve) of period $2\pi$ smoothly turn into Al'tshuler-Aranov-Spivak oscillations (black/dark grey curve) of period $\pi$ as the disorder strength is increased.}
\label{fig:ABtoAAS}
\end{figure}
\begin{figure}[h!]
\includegraphics[width=\linewidth]{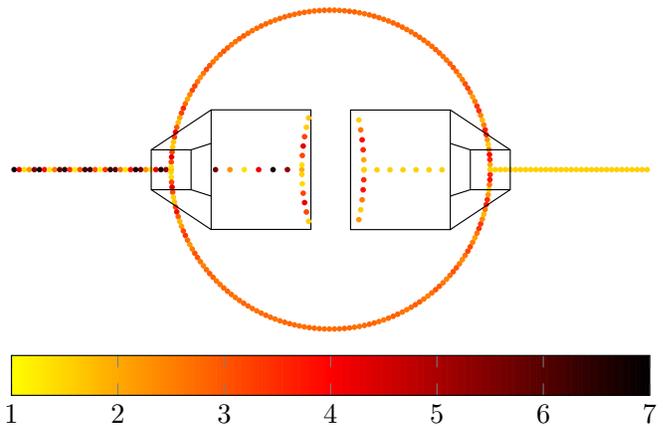}
\caption{Disorder averaged density of atoms at each site of the system at $\Phi = \pi/2$ in a noninteracting situation with $\mu/E_\delta = 0.2$ and $N_R = 200$ sites. We considered 20000 realisations of a gaussian correlated disorder taking random values in $[-0.0119, 0.0119]$ with correlation length $\sigma = 20 \delta$. Systematic interferences between particles take place at the entrance and exit junctions of the ring.}
\label{fig:dens} 
\end{figure}

As is shown in \textsc{Fig.}~\ref{fig:ABtoAAS}, Aharonov-Bohm oscillations are washed out by the ensemble average, giving rise to Al'tshuler-Aronov-Spivak oscillations of period $\pi$. \textsc{Fig.}~\ref{fig:dens} shows the disorder-averaged density of atoms on each site. It illustrates robust interferences that take place near the entrance and exit junctions of the ring. They arise because reflected particles interfere with the injected current.  Deep inside the ring as well as in the downstream region, on the other hand, a homogeneous mean density is encountered.

\subsection{Competition between disorder and interaction effects}
\begin{figure}
\includegraphics[width=1.05\linewidth, clip=true]{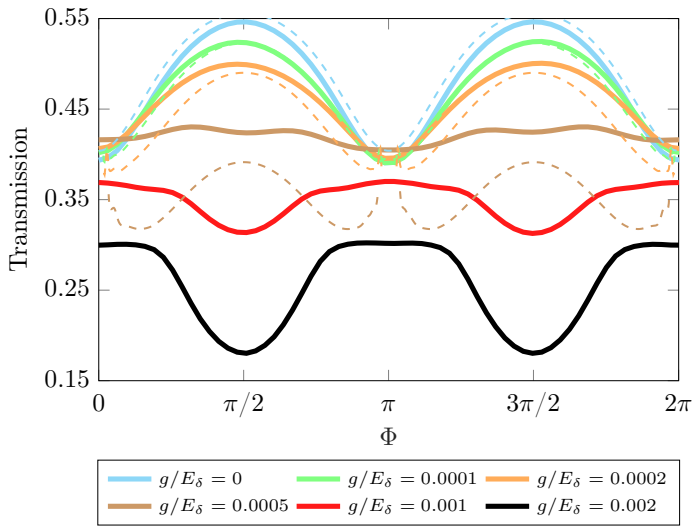}
\caption{Gross-Pitaevskii simulations showing the transmission versus $\Phi$ for different interaction strengths $g$ (increasing $g$ with increasing greyscale) with $\mu/E_\delta = 0.2$, $N_R=200$ sites and $\sqrt{\mathcal{N}} \vert \kappa \vert/E_\delta = 1$, which yields a density $\delta\rho^\varnothing = 2.77$. We considered 20000 realisations of a gaussian correlated disorder taking random values in $[-0.0119,0.0119]$ with correlation length $\sigma=20\delta$. The interaction gives first rise to a flattening and then to a reversal of the curve: the maxima at $\Phi = \pi/2$ and $3\pi/2$ become minima and the minima at $\Phi=0$ and $\pi$ become maxima. The predictions of our diagrammatic theory are represented in dashed lines of the same colour as the corresponding Gross-Pitaevskii simulations. They exhibit a good agreement with the latter for low values of the interaction strength. As soon as $g$ increases, quadratic corrections become more important and the predictions of a linear theory become less reliable.}
\label{fig:GPexp}
\end{figure}

\textsc{Fig.}~\ref{fig:GPexp} shows the results of a Gross-Pitaevskii simulation for different values of the interaction strength $g$ showing the transmission as a function of $\Phi$.
We see that from $g/E_\delta = 0$ to $g/E_\delta = 0.0005$, the presence of interaction gives rise to a flattening of the oscillations by reducing the amplitude. However, for $g/E_\delta = 0.001$ and $g/E_\delta = 0.002$, we observe an inversion of Al'tshuler-Aronov-Spivak oscillations. This is in qualitative agreement with coherent backscattering inversion \cite{Hartung2008PRL}. The minima of transmission at $\Phi=0$ and $\pi$ (corresponding to maxima of reflection in a coherent backscattering scenario) become maxima and the two maxima located around $\Phi = \pi/2$ and $3\pi/2$ become minima. We expect that those results remain quantitatively the same in a large range of parameters defining the disorder potential, provided the latter is sufficiently strong to fully randomize the phase factors $e^{i\pu}$ and $e^{i\pd}$ and at the same time sufficiently weak and smooth so that it does not induce (partial or total) reflections within each arm of the ring.

We also plotted on this figure the predictions of our first-order in $g$ diagrammatic theory which will be developed in Sec.~\ref{sec:diagth}. The agreement we find between the two curves is reasonably good for weak interaction strength $g/E_\delta \lesssim 0.0002$. At stronger interactions, significant deviations occur due to quadratic corrections becoming more important. 

\begin{figure}[h!]
\begin{center}
\includegraphics[width=\linewidth, clip=true]{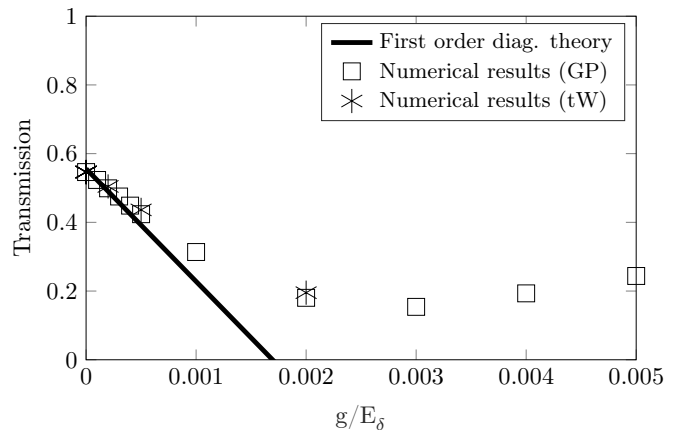}
\end{center} 
\caption{Comparison between the first-order diagrammatic theory predictions and the results of numerical Gross-Pitaevskii and truncated Wigner simulations for the transmission as a function of the interaction strength $g$ at $\Phi = \pi/2$, with $\mu/E_\delta = 0.2$ and $N_R=200$ sites. We considered 20000 realisations of a gaussian correlated disorder taking random values in $[-0.0119,0.0119]$ with correlation length $\sigma=20\delta$. Good agreement is found for weak interaction strength where the transmission decreases approximately linearly with $g$.}
\label{fig:compGPtWDiag}
\end{figure}

Another comparison with the predictions of our analytical diagrammatic theory is shown in \textsc{Fig.}~\ref{fig:compGPtWDiag} where we plot the transmission at $\Phi=\pi/2$ at a finite value of the interaction strength. We find that for small values of $g$, the transmission decreases linearly with $g$. The initial decrease of the transmission with $g$ is reasonably well predicted by our diagrammatic theory at about $g\simeq 10^{-3}E_\delta$ beyond which quadratic corrections in $g$ become important. A maximal inversion of Al'tshuler-Aronov-Spivak oscillations is reached at about $\Phi \simeq 0.003 E_\delta$.

\begin{figure}[h!]
\begin{center}
\includegraphics[width=\linewidth, clip=true]{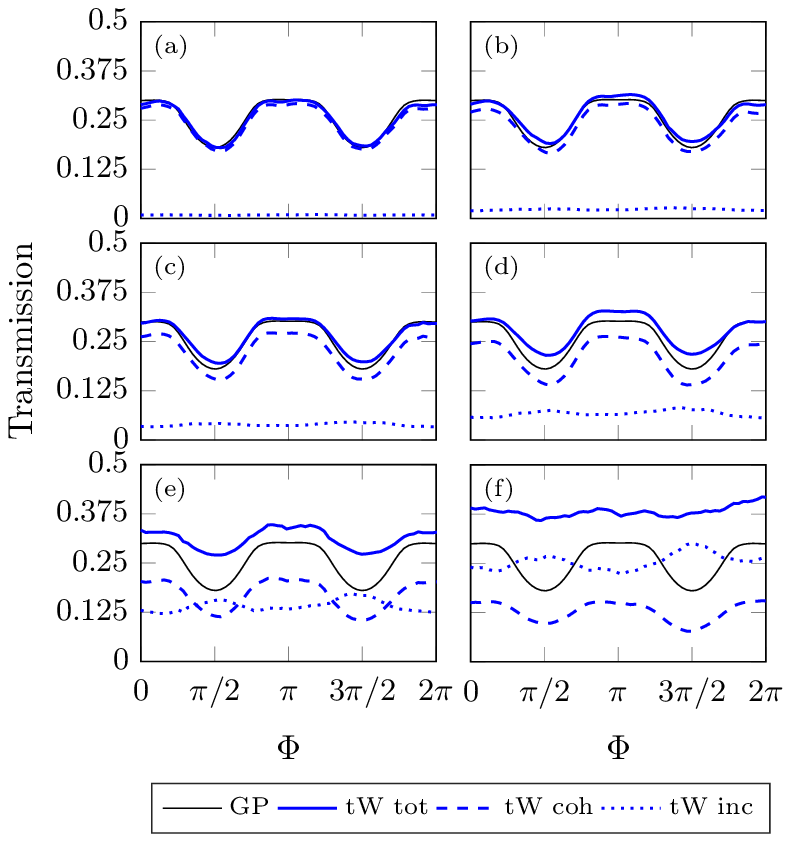}
\end{center} 
\caption{Truncated Wigner simulation showing the transmission versus $\Phi$ for different interaction strengths $g$ and $\rho^\varnothing$, the product $\delta g\rho^\varnothing/E_\delta \simeq 0.0055$ being kept constant all the way. In the Gross-Pitaevskii simulations (thin black line), this corresponds to a regime where the inversion of Al'tshuler-Spivak oscillations is fully developed. The simulation parameters are $\mu/E_\delta = 0.2$ and $N_R=200$ sites. We considered 20000 realisations of a gaussian correlated disorder taking random values in $[-0.0119,0.0119]$ with correlation length $\sigma=20\delta$, each of which done with an average of 100 realisations over the initial conditions. The values of both $g/E_\delta$ and $\delta\rho^\varnothing$ are $g/E_\delta=0.0002$ and $\delta\rho^\varnothing\simeq 27.77$ for (a), $g/E_\delta=0.001$ and $\delta\rho^\varnothing\simeq 5.55$ for (b), $g/E_\delta=0.002$ and $\delta\rho^\varnothing\simeq 2.77$ for (c), $g/E_\delta=0.004$ and $\delta\rho^\varnothing\simeq 1.38$ for (d), $g/E_\delta=0.01$ and $\delta\rho^\varnothing\simeq 0.55$ for (e), $g/E_\delta=0.02$ and $\delta\rho^\varnothing\simeq 0.27$ for (f).}
\label{fig:tWAAS}
\end{figure}

Finally, we performed truncated Wigner simulations in a regime where the inversion of Al'tshuler-Aronov-Spivak oscillations is fully developed, namely for $g/E_\delta = 0.002$. We have therefore performed six sets of simulations, with different values for both the density $\rho^\varnothing$ and $g$, their product being kept constant and equal to $\delta g\rho^\varnothing/E_\delta \simeq 0.0055$. The results of these simulations are shown in \textsc{Fig.}~\ref{fig:tWAAS}. We clearly see that truncated Wigner simulations predict a flattening of inverted Al'tshuler-Aronov-Spivak transmission profile, corresponding to a complete dephasing of quantum interference effects.

We furthermore see that the inverted Al'tshuler-Aronov-Spivak oscillation structure arises entirely due to a coherent contribution, thereby confirming that the inversion of the central minimum has the same origin as coherent backscattering inversion. The coherent part formally still exhibits this inverted structure, but is hidden behind the incoherent contribution which has become very large, indicating the presence of dephasing for strong interaction. 

\section{Diagrammatic theory for the coherent contributions to the Al'tshuler-Aronov-Spivak oscillations \label{sec:diagth}}

\subsection{Formal solution of the Gross-Pitaevskii equation}
The starting point of an analytical diagrammatic theory on mean-field level is the Gross-Pitaevskii equation of the discretised system, which is given by Eq.\ \eqref{eq:SchrSourterm}. This equation is formally solved by
the time-dependent scattering wavefunction
\begin{equation}
  \psi_\alpha(t)=\sum_{\alpha'} G(\alpha,\alpha',\mu) S_{\alpha'}(t)\,,
\end{equation}
where we used the linear Green function for the system without
interaction. The source term as well as the nonlinear interaction
term are contained in
\begin{equation}
  S_\alpha(t) = g_\alpha \vert\psi_\alpha(t)\vert^2 \psi_\alpha(t)
    + \kappa(t) \delta_{\alpha,\alpha_{\mathcal{S}}} \sqrt{\mathcal{N}}.
\end{equation}
In the limit of long times, when we have reached a stationary state,
the Gross-Pitaevskii equation transforms into a self-consistent
equation,
\begin{equation}
  \psi_\alpha
  = \sqrt{\mathcal{N}} \kappa(t) G(\alpha,\alpha_S,\mu) 
  + \sum_{\alpha'} G(\alpha,\alpha',\mu) g_{\alpha'}
  \vert\psi_{\alpha'}\vert^2 \psi_{\alpha'}\,,
  \label{eq:GPE_solution_selfconsistent}
\end{equation}
which marks the starting point for a perturbation theory in the small
interaction parameter $g$. The zeroth order in this expansion marks
the noninteracting case.

\subsection{The noninteracting case}

In this section we present the calculation of the Green function for
the noninteracting case and for a fixed disorder configuration.  This
calculation is based on its representation by a sum over all paths
linking two sites (see Appendix \ref{sec:GFprod}),
\begin{equation}
  G(\alpha,\alpha',\mu)
  = \dfrac{1}{i E_\delta \sin(k \delta)}
  \sum_\gamma A_\gamma e^{iS_\gamma/\hbar}.
  \label{eq:GF_as_path_sum}
\end{equation}
To visualise this, it is useful to think of the system as a
quantum graph consisting of two semi-infinite waveguides, which are on
opposite sides attached to the ring structure via two
junctions. Depending on the locations $\alpha,\alpha'$ at the beginning and the
end of $\gamma$, the path might visit one or both of the junctions and
fully explore the branches of the ring multiple times. Every path
$\gamma$ in the coherent sum in Eq.~(\ref{eq:GF_as_path_sum}) may
contain an arbitrarily long sequence of such alternating visits of
junctions and explorations of branches.

The phase factor $\exp(i/\hbar S_\gamma)$ contains the accumulated
phase of the repeated exploration of the branches. A single traversal
of a branch contributes a random but fixed phase $\Phi_u,\Phi_d$ due
to the disorder potential. Additionally, we get another phase
contribution $\pm\Phi/2$, whose magnitude depends on the flux enclosed
by the ring, and its sign encodes whether the flux is encircled in
counterclockwise or clockwise direction. The crossing of a
junction is treated within a scattering approach and yields a
multiplicative contribution to the amplitude $A_\gamma$, either a
reflection $r$ or a transmission amplitude $t$, depending on the
geometry of the path $\gamma$ before and after the junction.

To keep track of the contributions of a single path to the coherent
sum in Eq.~(\ref{eq:GF_as_path_sum}), we establish the following
diagrammatical representations,
\begin{equation}
  \includegraphics[width=0.8\linewidth]{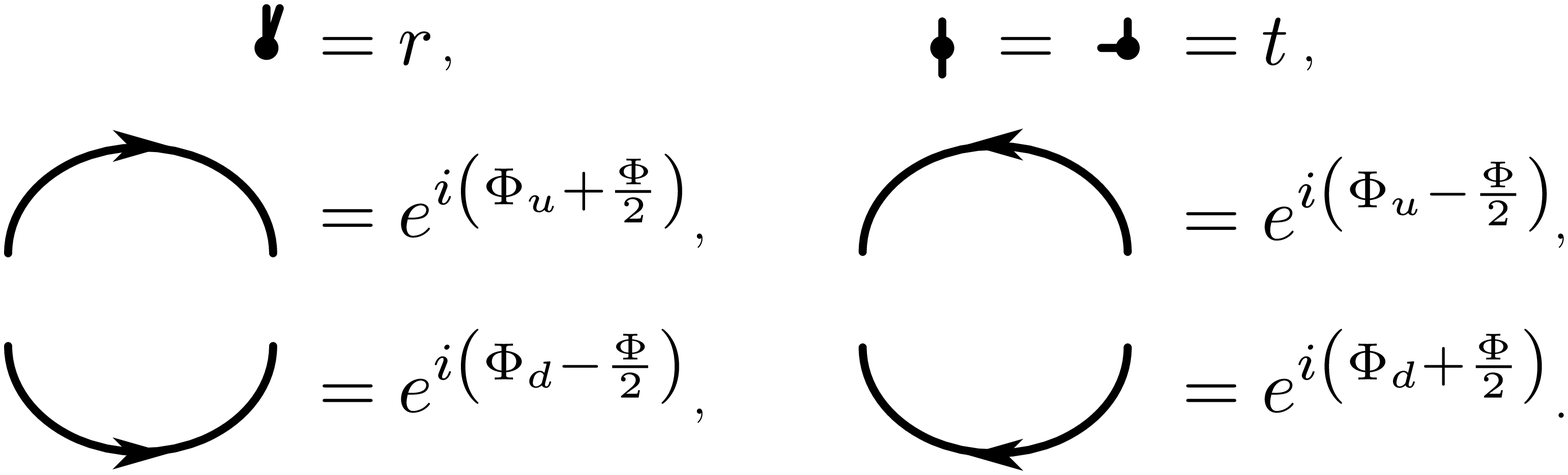}
  \label{eq:basic_defs}
\end{equation}
Every path can be visualised as a sequence of the above building blocks,
where the sequence alternates between the scattering process at a
junction [first line in Eq.~(\ref{eq:basic_defs})] and the exploration
of one of the branches, either in clockwise or counterclockwise direction
[second and third line in Eq.~(\ref{eq:basic_defs})].

Our aim is to perform the coherent sum over all paths inherent in the
noninteracting Green function. To do so, we first focus on the
sequential part oscillating between the two junctions. We group these
sequences depending on which junction they start and end, and how they
approach and leave the limiting junctions before and after the
sequence. This group of sequences with identical limiting conditions
is then resummed and represented by a single new diagram in the colour
orange (dark grey) and an arrow representing the common
initial and final behavior. For instance, the subsequent summation,
\begin{equation}
  \includegraphics[width=0.6\linewidth]{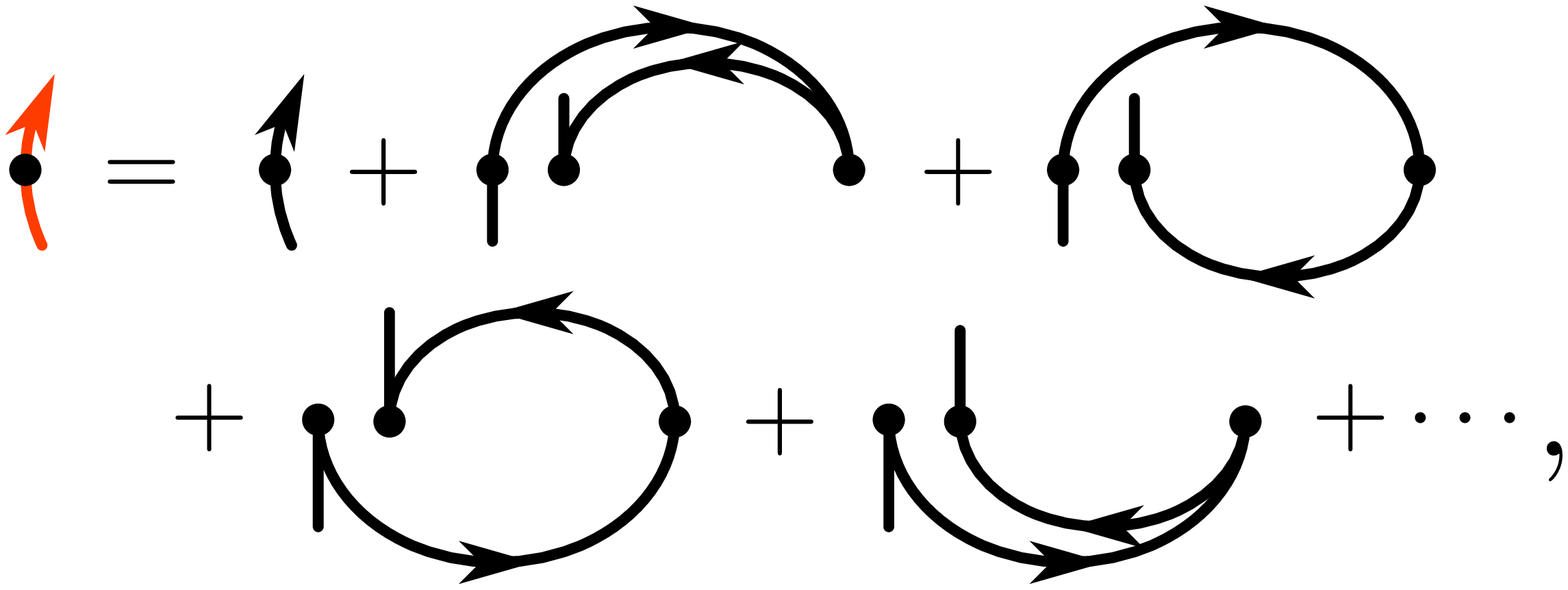}
  \label{eq:LD_LU-explicit_sum}
\end{equation}
represents the sum of all trajectories that approach the left
junction from the lower branch, have an arbitrary number of
explorations of upper and lower branches of the Aharonov-Bohm ring,
and then finally leave to the upper branch. The boundary conditions,
as well as the junction they refer to, are encoded in the direction
and the curvature of the orange (dark grey) arrow. This way of thinking defines a
whole set of resummed diagrams,
\begin{equation}
  \includegraphics[width=0.8\linewidth]{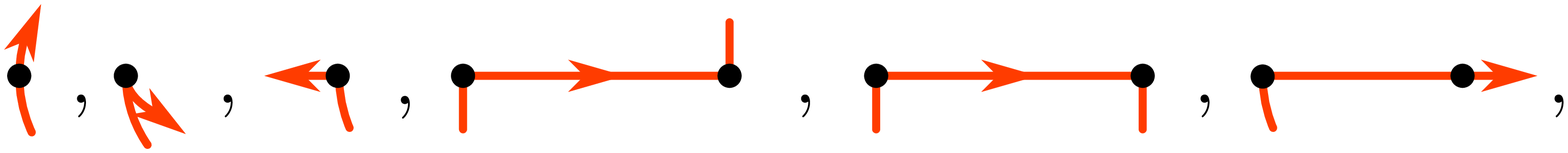}
  \label{eq:LD_any-presentation}
\end{equation}
as well as rotated and mirrored versions of the above diagrams. The
selected resummed diagrams in Eq.~(\ref{eq:LD_any-presentation}) all represent resummations of
paths that approach the left junction from the lower branch at the
beginning. The first three end at the left junction, while the last
three end at the right one. From there, the diagrams either leave to
the upper or the lower branch of the ring, or exit the ring to the
attached waveguide.

The calculation of these resummed diagrams utilises self-consistent
equations. Those are constructed by the observation that every path
contained in the sums of the diagrams
Eq.~(\ref{eq:LD_any-presentation}) can be extended to a longer path by
successively adding two branches and two junctions. By appropriately
adjusting the boundary conditions, these longer paths have to be an
element of a resummed diagram. For instance, we find
\begin{equation}
  \includegraphics[width=0.85\linewidth]{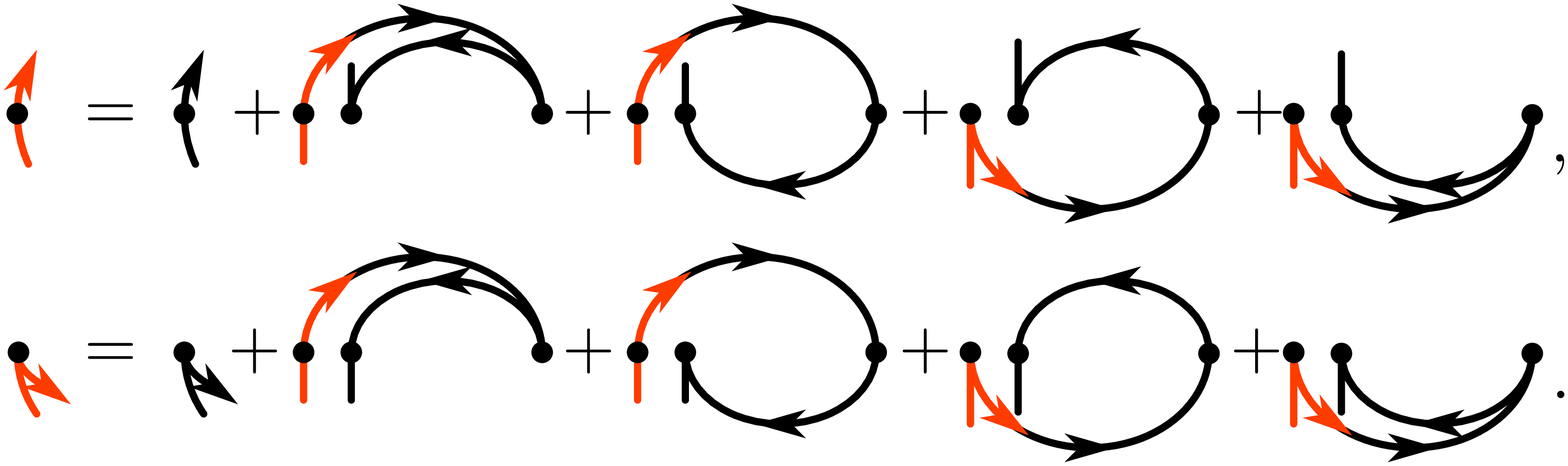}
\end{equation}
We can rewrite these equations in a diagrammatic matrix-vector version,
\begin{equation}
  \includegraphics[width=0.85\linewidth]{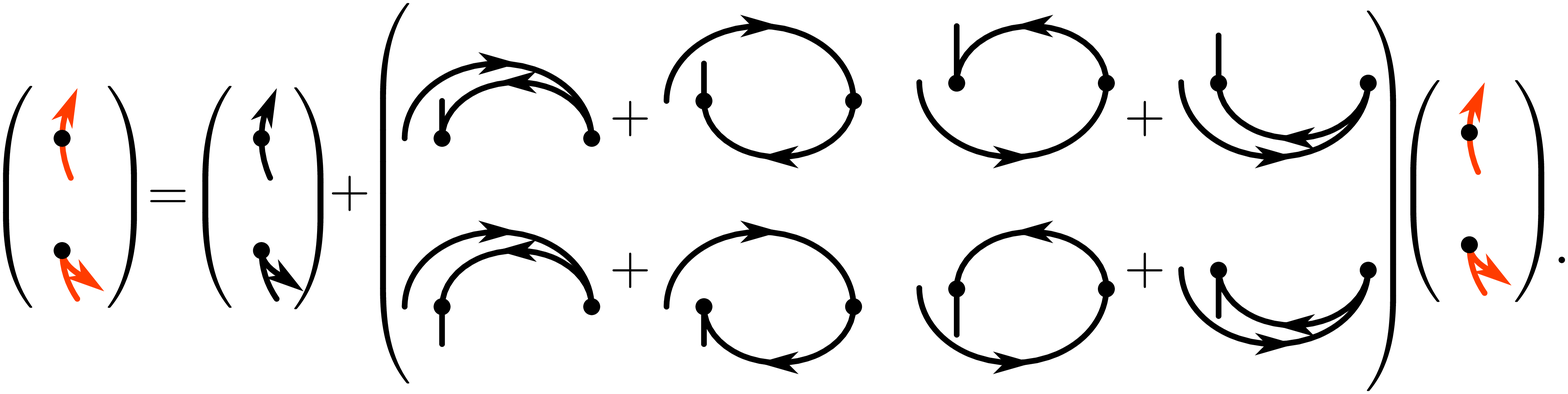}
  \label{eq:selfcons_matrix}
\end{equation}
To avoid a repeated use of the same diagrams in the subsequent
calculations, let us denote the two-dimensional vector with the
resummed diagrams, which appears on both sides of
Eq.~(\ref{eq:selfcons_matrix}), by $\vec{w}$, and the first vector at
the right hand side of Eq.~(\ref{eq:selfcons_matrix}) by $\vec{v}$. The matrix
will be denoted by $A$. Then by applying the rules in
Eq.~(\ref{eq:basic_defs}) the diagrams translate to
$\vec{v}=(t,r)^{\intercal}$ and
\begin{equation}
  \begin{aligned}
    &A=\\
    &\begin{pmatrix}
      r^2  e^{2i \pu}
      + t^2 e^{i \left( \pu + \pd + \p \right)}
      &
      rt e^{i\left( \pu + \pd - \p \right)}
      +
      rt e^{2i\pd}
      \\
      rt e^{2i\pu}
      +rt e^{i \left( \pu + \pd + \p \right)}
      &
      t^2e^{i \left( \pu + \pd - \p \right)}
      +r^2 e^{2i\pd}
    \end{pmatrix}.
  \end{aligned}
\end{equation}
The self-consistent Eq.~(\ref{eq:selfcons_matrix}) translates to
$\vec{w}=\vec{v}+A\vec{w}$ whose solution $\vec{w}$ is found by a
simple matrix inversion, $\vec{w}=(1-A)^{-1}\vec{v}$. We obtain
\begin{equation}
  \begin{aligned}
    &\raisebox{-0.5\height}{
      \includegraphics[width=0.075\linewidth]{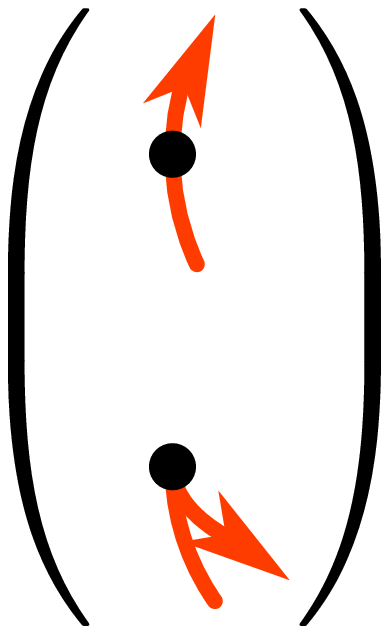}
    }=
    \frac{1}{D(\pu,\pd,\p)}
    \begin{pmatrix}
      t\left(1 + \left( r^2 - t^2 \right)
        e^{i \left(\pu+\pd-\p \right)}\right)\\
      r \left(1-\left(r^2-t^2\right) e^{2i\pu}\right)
    \end{pmatrix},
  \end{aligned}
  \label{eq:LD_Lx_result}
\end{equation}
where the common denominator is the determinant of the matrix $(1-A)$,
\begin{equation}
  \begin{aligned}
    &D(\pu,\pd,\p) = \det(1-A)\\
    &= 1 -  r^2 \left( e^{2 i\pu} + e^{2 i\pd} \right)
    -  2t^2 e^{i(\pu+\pd)} \cos(\p)\\
    &\quad + \left(r^2-t^2\right)^2 e^{2i(\pu+\pd)}.
  \end{aligned}
  \label{eq:general_denominator}
\end{equation}
In principle, the other diagrams in Eq.~(\ref{eq:LD_any-presentation})
can be calculated in a similar manner, but this is not
necessary. Having one pair of resummed diagrams, it is possible to
derive all the other diagrams in Eq.~(\ref{eq:LD_any-presentation}),
as well as their mirrored and rotated versions. To construct a new
path from a given path with a different behavior at the initial or
final junction, it is sufficient to elongate the original path by one
or two branches, thereby incorporating the desired behaviour at the new
endings. This leads to a set of diagrammatic identities, which we
present in the Appendix \ref{sec:appendix_diagrammatic}.

We are now ready to write a diagrammatic expression for the linear
Green function $G(\alpha,\alpha',E)$. For instance, if the sites
$\alpha$ and $\alpha'$ are in the upper branch of the ring, we have
\begin{equation}
  \includegraphics[width=0.9\linewidth]{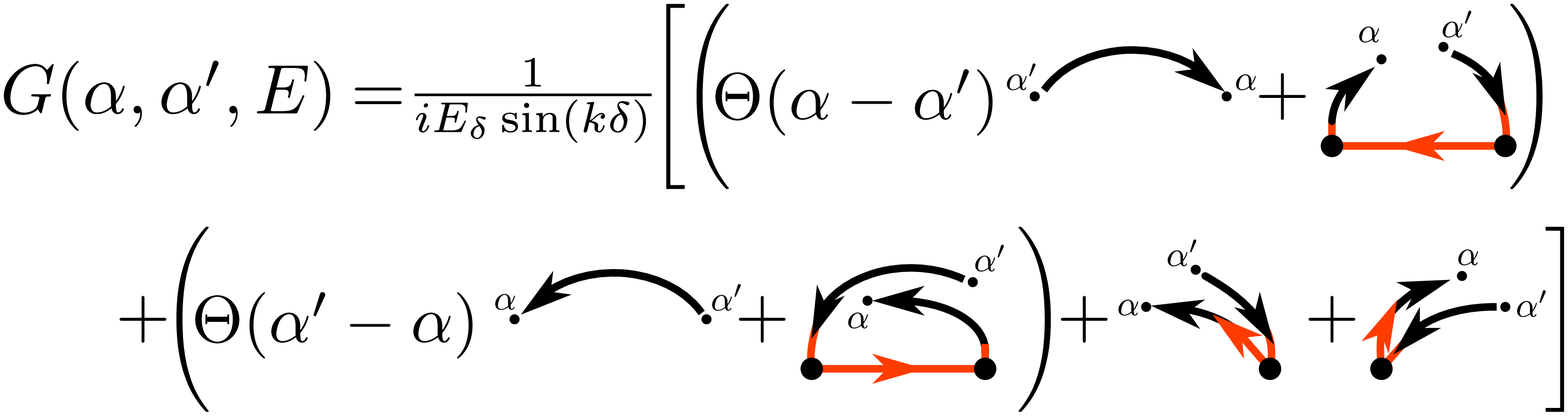}
  \label{eq:linear_GF}
\end{equation}
where the Heaviside function $\Theta(\cdot)$ distinguishes the cases where
$\alpha$ is left ($\alpha<\alpha'$) or right of $\alpha'$
($\alpha>\alpha'$). Note that as these representations include the
partial exploration of a branch, details of the disorder in the single
branches are needed to obtain the correct phase. However, as we are
interested in transport through the ring, we will see that it is
sufficient to consider only paths that terminate at junctions, i.e.,
which only contain full explorations of the branches of the ring.

For the calculation of the full noninteracting reflection amplitudes
$R^{(0)}$ we need to consider a Green function that starts and ends
at a site in the left waveguide. This requires to sum over all paths,
that start and end in that waveguide. Utilising the resummed
diagrams, we obtain, up to a constant phase,
\begin{equation}
  \begin{aligned}
   &\includegraphics[width=0.7\linewidth]{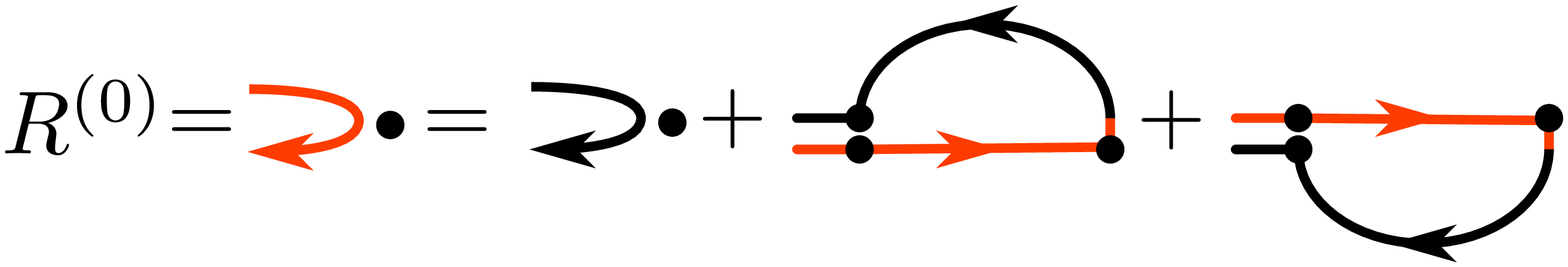}\\
   &=
   r + \frac{t^2}{D(\pu,\pd,\p)}\Big[
   -2\left(r-t\right)\left(r^2-t^2\right)
    e^{i 2\left(\pu+\pd\right)}\\
    &\qquad +    r \left(e^{i 2\pu}+e^{i 2\pd}\right)
    + 2 t e^{i\left(\pu+\pd\right)}\cos(\p)\Big]\,.
  \end{aligned}
  \label{eq:refl_nonint}
\end{equation}
In a similar way, we can also calculate the full noninteracting
transmission amplitude $T^{(0)}$,
\begin{equation}
  \begin{aligned}
    & \includegraphics[width=0.7\linewidth]{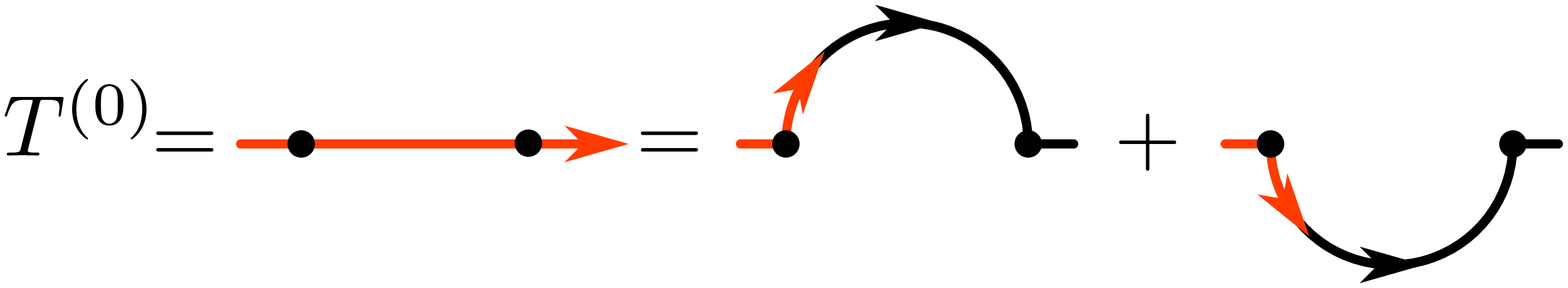}\\
    & =\frac{t^2}{D(\pu,\pd,\p)}\Big[
    e^{i\left(\pu+\frac{\p}{2}\right)}+
    e^{i\left(\pd-\frac{\p}{2}\right)}\\
    &\qquad
    -\left(r-t\right)^2e^{i\left(\pu+\pd\right)}
    \left(
      e^{i\left(\pu-\frac{\p}{2}\right)}+
      e^{i\left(\pd+\frac{\p}{2}\right)}
    \right)\Big]\,.
  \end{aligned}
  \label{eq:trans_nonint}
\end{equation}
To obtain from these results the disorder-averaged noninteracting
reflection and transmission probability, we have to take the modulus
square of the amplitudes in Eqs.~(\ref{eq:refl_nonint}),
(\ref{eq:trans_nonint}), and average them over all disorder phases
$\pu,\pd\in[0,2\pi]$;
\begin{align}
  \overline{\left|R^{(0)}\right|^2}(\p)
  &=
  \frac{1}{(2\pi)^2}
  \int_0^{2\pi} d \pu
  \int_0^{2\pi} d \pd |R^{0}(\pu,\pd,\p)|^2,
  \\
  \overline{\left|T^{(0)}\right|^2}(\p)
    &=\frac{1}{(2\pi)^2}
    \int_0^{2\pi} d \pu
    \int_0^{2\pi} d \pd |T^{(0)}(\pu,\pd,\p)|^2\,.
\end{align}
We performed the calculation of those integrals numerically with a Monte-Carlo approach. The result of this calculation agrees very well with the numerical simulation, as shown in \textsc{Fig.}~\ref{fig:GPexp}.

\subsection{The interacting case}
For the diagrammatic representation of the full interacting solution
$\psi_\alpha$, Eq.~(\ref{eq:GPE_solution_selfconsistent}), we introduce a similar diagrammatic theory as for the
noninteracting case. Our aim is thus to distinguish sets of
``interacting trajectories'' according to whether their final point
$\alpha$ in the index of $\psi_\alpha$ is in the upper or the lower branch
of the Aharonov-Bohm ring, and whether $\alpha$ is approached from its
left or right side. The wave function $\psi_\alpha$ is then
written as sum of these diagrams, \emph{e.g.}, for $\alpha$ in the upper part
of the ring,
\begin{equation}
  \includegraphics[width=0.6\linewidth]{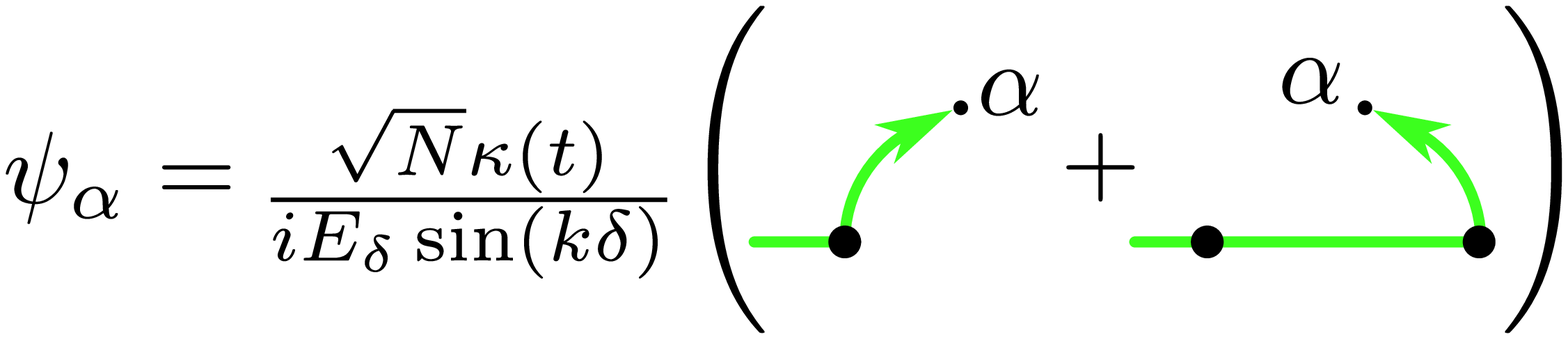}
  \label{eq:GPE_wavefunction}
\end{equation}
where the green(light grey)-coloured diagrams have to be understood in view of the
self-consistent equation,
Eq.~(\ref{eq:GPE_solution_selfconsistent}). On the right hand side of that
equation, the final site $\alpha$ is approached by paths inherent in
the linear Green function $G(\alpha,\alpha',E)$. The two green(light grey)-coloured
diagrams in Eq.~(\ref{eq:GPE_wavefunction}) account for those two possibilities, namely that the final site is approached in clockwise or counterclockwise
direction.

In the Gross-Pitaevskii Eq.~(\ref{eq:GPE_solution_selfconsistent}) we
represent, in close analogy with the nonlinear diagrammatic theory for 2D billiards \cite{Hartmann2012AP}, by a box the intermediate site $\alpha'$ where the
nonlinear wave interacts with itself through the density
$|\Psi_{\alpha'}|^2$. Complex conjugation is diagrammatically
expressed by replacing solid lines with dashed ones. Then
Eq.~(\ref{eq:GPE_solution_selfconsistent}) transforms into a set of
diagrammatic equations. One of them is
\begin{widetext}
  \begin{equation}
    \includegraphics[width=\linewidth]{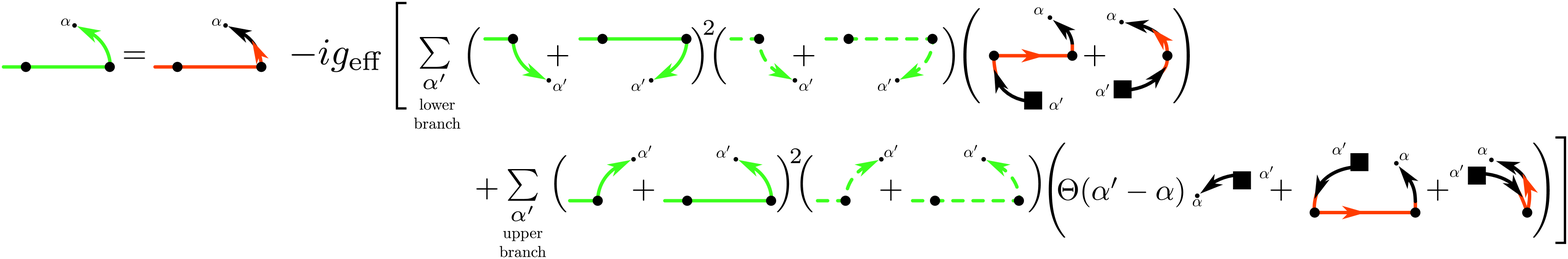}
    \label{eq:GPE_diagrammatic}
  \end{equation}
\end{widetext}
where
\begin{equation}
  \geff = \frac{g \mathcal{N} \left|\kappa(t)\right|^2}
  {[E_\delta \sin(k \delta)]^3}
\end{equation}
denotes the effective interaction strength. By subsequently inserting the left hand side into the right hand side,
Eq.~(\ref{eq:GPE_diagrammatic}) and its relatives can be used for a
perturbative calculation in the small effective interaction strength
$\geff$.

To obtain the diagrams for the full reflection (and transmission)
amplitude for the interacting case, we have to send the site $\alpha$
in Eq.~(\ref{eq:GPE_wavefunction}) to the left (resp.
right) junction site and add a transmission event to leave the ring.
We obtain
\begin{equation}
  \includegraphics[width=0.825\linewidth]{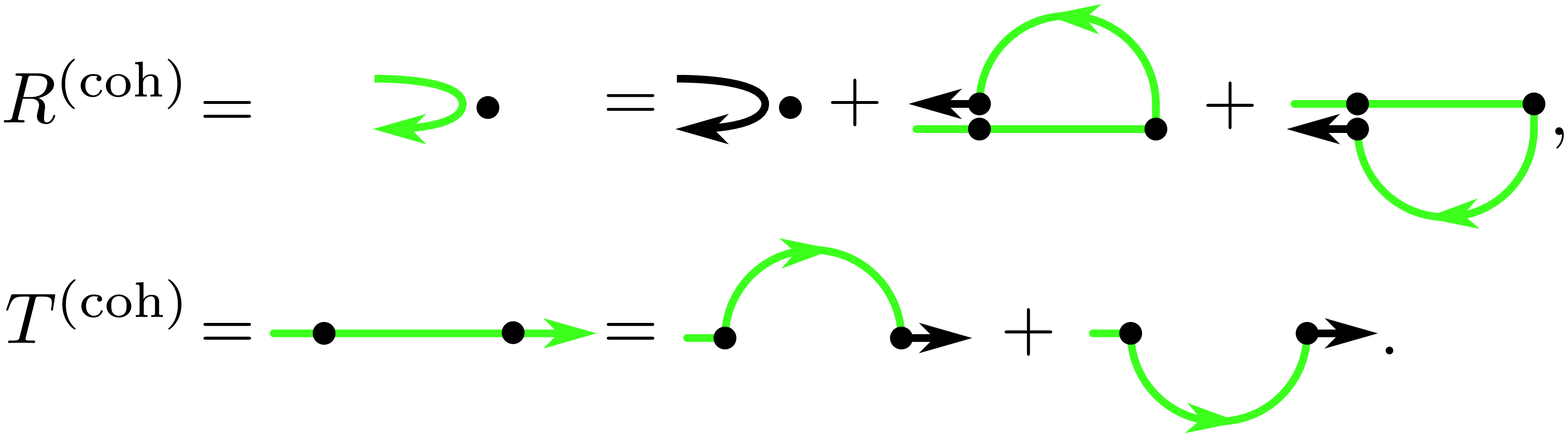}
  \label{eq:full_refl_trans}
\end{equation}
The exact calculation of these expressions in first and higher orders
of the effective interaction strength $\geff$, however, requires details
of the specific disorder at the single sites: To perform the summation
over all sites in the ring in
Eqs.~(\ref{eq:GPE_solution_selfconsistent}),
(\ref{eq:GPE_diagrammatic}) we have to find the accumulated phase from
a partial exploration of one of the branches. However, to arrive at the desired
averaged reflection and transmission probability, the modulus square of
Eq.~(\ref{eq:full_refl_trans}) has to be averaged over disorder
configurations. The only contributions
surviving the averaging process are those where the accumulated phase
of a partial exploration is, with the help of other paths, either
extended to the phase of a full exploration, or compensated by
other partial explorations with an opposite sign in front of their
phases. Consequently, this is only the case for path constellations
where the nonlinearity event displays one of the following
structures,
\begin{equation}
  \includegraphics[width=0.5\linewidth]{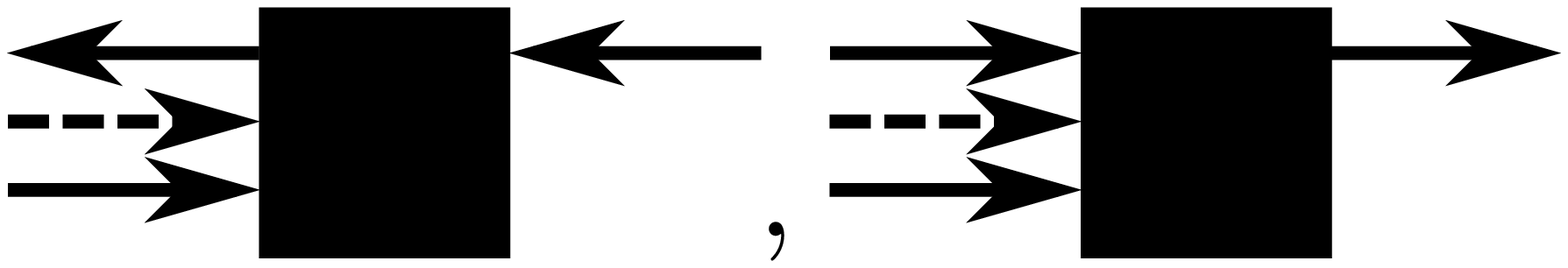}
\end{equation}
as well as the mirrored and/or complex conjugated versions of
those. Contrarily, structures involving
\begin{equation}
  \includegraphics[width=0.66\linewidth]{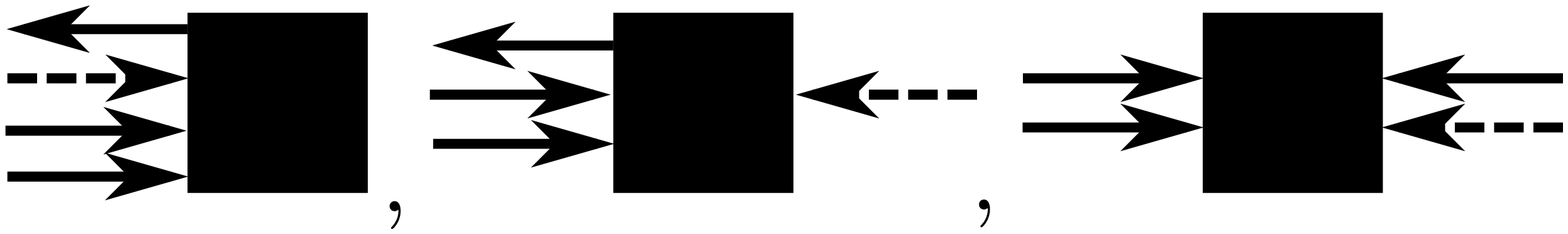}
  \label{eq:forbidden_structures}
\end{equation}
produce an uncompensated accumulated phase which does not survive the
disorder average.  \footnote{In principle, to compensate the phase of
  one of the forbidden structures, a second nonlinearity event, which
  moves together with the first would be needed. However, this
  requires to reduce two summations over the positions of the
  nonlinearity events to a single one. Their contribution is thus
  suppressed by a factor of the order of the inverse number of sites
  compared to those including only allowed structures.}

The above arguments motivate excluding the structures in
Eq.~(\ref{eq:forbidden_structures}) from our diagrammatic theory. This
allows to perform calculations without the need to know exact details
of the specific disorder configuration, while still producing accurate
results for the disorder-averaged reflection and transmission
probabilities. Eq.~(\ref{eq:GPE_diagrammatic}) is substituted by
\begin{widetext}
  \begin{equation}
    \includegraphics[width=0.9\linewidth]{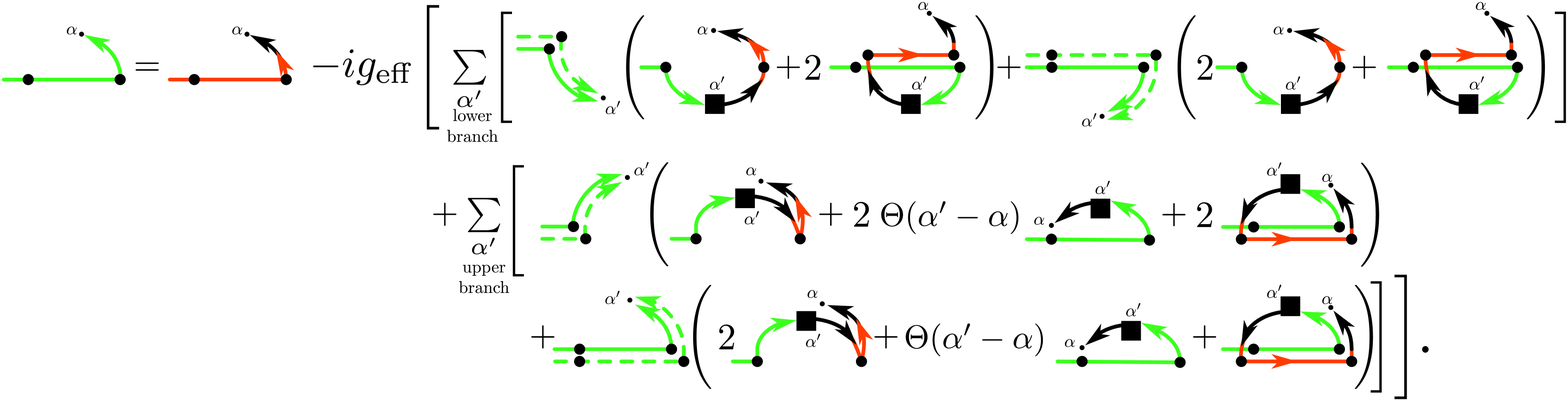}
    \label{eq:GPE_effective}
  \end{equation}
\end{widetext}
The additional factor 2 appearing in the above terms is a
combinatorical factor, reflecting the two possibilities to build the
corresponding contributions from the single diagrammatic parts in
Eq.~(\ref{eq:GPE_diagrammatic}).

The first order of the perturbative expansion of
Eq.~(\ref{eq:GPE_effective}) in orders of $\geff$ is obtained by
replacing green (light grey) diagrams by orange (dark grey) ones with a subsequent partial
exploration of a branch,
\begin{equation}
  \includegraphics[width=0.6\linewidth]{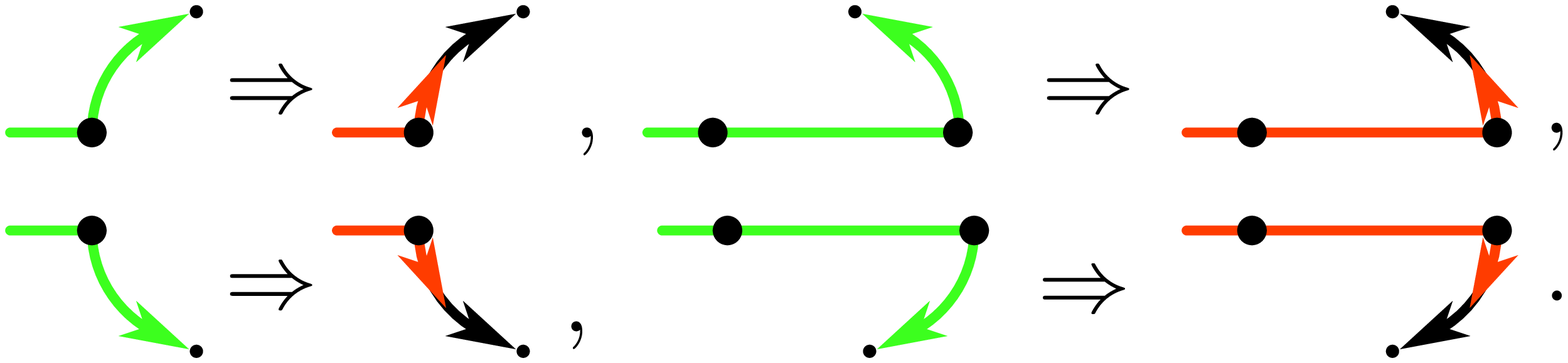}
\end{equation}
Within this approximation, the terms in the summation over the
intermediate site $\alpha'$ in Eq.~(\ref{eq:GPE_effective}) are
independent on $\alpha'$ and the summation is easily performed,
yielding an additional factor $N_R$, the number of sites in a single
branch of the ring.

By further utilising diagrammatic equalities for resummed paths, we
obtain that the effective corrections in first order in $\geff$ to the
reflection and transmission amplitudes are diagrammatically given by
\begin{widetext}
  \begin{equation}
    \includegraphics[width=0.8\linewidth]{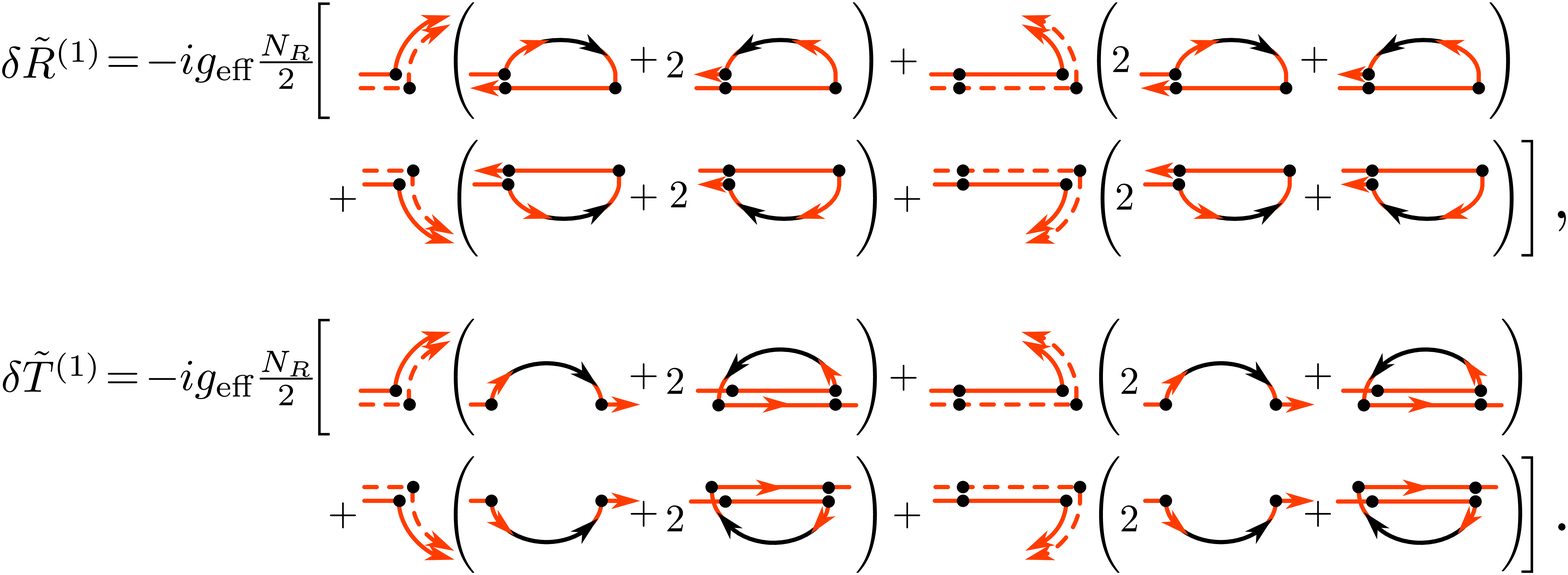}
  \end{equation}
\end{widetext}
By using the calculated analytical expressions for the appearing
resummed diagrams, we can obtain analytical expressions for these
contributions. These expressions can then be used to numerically
calculate the first-order correction according to
\begin{align}
  &\delta \overline{|R^{(1)}|^2}(\p)
  =\overline{|R^{(1)}|^2}(\p) - \overline{|R^{(0)}|^2}(\p)
  \\
  &=
  \iint_0^{2\pi} \frac{d \pu d \pd}{(2\pi)^2}2
  \mathrm{Re}\left[
      R^{(0)*}(\pu,\pd,\p)
      \delta \tilde{R}^{(1)}(\pu,\pd,\p)
    \right]\nonumber
\end{align}
\begin{align}
  &\delta \overline{|T^{(1)}|^2}(\p)
  =\overline{|T^{(1)}|^2}(\p) - \overline{|R^{(0)}|^2}(\p)
  \\
  &=
  \iint_0^{2\pi} \frac{d \pu d \pd}{(2\pi)^2}2
  \mathrm{Re}\left[
      T^{(0)*}(\pu,\pd,\p)
      \delta \tilde{T}^{(1)}(\pu,\pd,\p)
    \right]
    \nonumber
\end{align}
Like for the noninteracting case, we carried out the above averaging through a numerical integration using a Monte-Carlo method. Nonzero first-order corrections are obtained essentially owing to the complexity of the reflection and transmission matrix elements at a Y junction, according to Eqs.\ \eqref{eq:rwc} and \eqref{eq:twc}. Note that they would vanish for the choice of a widely "open" ring with $r=-1/3$ and $t=2/3$ (Eqs.\ \eqref{eq:rcl} and \eqref{eq:tcl}).

\section{Conclusions \label{sec:concl}}
In summary, we numerically studied the 1D transport of a Bose-Einstein condensate through a lead connected to a disordered ring-shaped geometry which is penetrated by an artificial gauge field. This particular setting is an ideal candidate for studying coherent backscattering and its interplay with atom-atom interaction beyond the mean-field approach.

The presence of very weak interaction is first accounted for in the framework of the mean-field Gross-Pitaevskii approximation. We observed an inversion of Al'tshuler-Aronov-Spivak oscillations, similarly to the inversion of coherent backscattering \cite{Hartung2008PRL,PhysRevLett.100.033902,Hartmann2012AP}. In contrast with Refs.~\cite{Hartung2008PRL,Hartmann2012AP}, a linear scaling of the inversion with $g$ is found, essentially due to the complexity of reflection and transmission amplitudes across junctions, as confirmed by nonlinear diagrammatic theory. It would imply that a further enhancement of Al'tshuler-Aronov-Spivak oscillations should be expected for weak negative $g$, provided a stationary scattering state could be realised in that case.

Truncated Wigner simulations show that this many-body effect is subject to dephasing, in qualitative agreement with \cite{Geiger2013}. This dephasing is phenomenologically very similar to the effect of the presence of a thermal bath (both for Aharonov-Bohm \cite{FerrierPRL2004,RischauPRB2017,RussoPRB2008} and Al'tshuler-Aronov-Spivak \cite{IyeSupMic2003,Yagi2012} oscillations), the simulation of which could be carried out by very similar methodology. 

More quantitatively, considering that the source injects atoms in the waveguide with speed of $v = 1$ mm/s, that the s-wave scattering length is $a_S = 5.313 \times 10^{-9}$ m for $^{87}$Rb and the confinement frequency is $\omega_\perp = 2\pi \times 1$ kHz, we find that the injected atomic density would correspond to the value $\delta\rho^\varnothing \simeq 0.0826$. From \textsc{Fig.}~\ref{fig:tWAAS}, we judge that this set of parameters would correspond to a situation where a possible inversion of Al'tshuler-Aronov-Spivak oscillations would be overshadowed by dephasing, thereby giving rise to a structureless (and dominantly incoherent) transmission profile as a function of $\Phi$. Other atomic species such as $^{39}$K, where the s-wave scattering length can be tuned to very small values \cite{Roati2008N} may be more suitable to realise an inversion of Al'tshuler-Aronov-Spivak oscillations in an experiment.

\begin{acknowledgments}
The authors want to thank Thomas Wellens and Tsampikos Kottos for fruitful discussions. The computational resources have been provided by the Consortium des Equipements de Calcul Intensif (CÉCI), funded by the F.R.S.-FNRS under Grant No. 2.5020.11. J.R. acknowledges financial support from Deutsche Forschungsgemeinschaft through Project No. Ri681/14-1 and GRK 1570.
\end{acknowledgments}

\appendix

\section{Derivation of the Green function \label{sec:GFprod}}
We can write the total wavefunction of the system as
\begin{equation}
\psi = P_{\mathcal{L}_L} \psi_{\mathcal{L}_L} + P_{\mathcal{R}} \psi_{\mathcal{R}} + P_{\mathcal{L}_R} \psi_{\mathcal{L}_R},
\end{equation}
with $P_X$ a projector onto either the left lead ($\mathcal{L}_L$), the ring ($\mathcal{R}$) or the right lead ($\mathcal{L}_R$) satisfying
\begin{equation}
\quad P_XP_Y = \delta_{XY} P_X  
\end{equation}
for all $X,Y = \mathcal{L}_L,\mathcal{R},\mathcal{L}_R$. Following a scattering matrix formalism, the wavefunction in the left and right lead can be cast into the form
\begin{align}
\psi_{\mathcal{L}_L} & = G_{\mathcal{L}_L} S \nonumber \\
& \hspace{0.25cm} + G_{\mathcal{L}_L} W_{\mathcal{L}_L\mathcal{R}} \left( G_\mathcal{R}^{-1} - \Sigma_{\mathcal{L}_L} - \Sigma_R \right)^{-1} W_{\mathcal{R}\mathcal{L}_L} G_{\mathcal{L}_L} S \\
\psi_{\mathcal{L}_R} & = G_{\mathcal{L}_R} W_{\mathcal{L}_R\mathcal{R}} \left( G_\mathcal{R}^{-1} - \Sigma_{\mathcal{L}_L} - \Sigma_{\mathcal{L}_R} \right)^{-1} W_{\mathcal{R}\mathcal{L}_L} G_{\mathcal{L}_L} S,
\end{align}
where we have introduced the source term $S$ and the free Green function
\begin{equation}
G_X = \left( \mu - H_X + i \epsilon \right)^{-1},
\end{equation}
with the notation $X = \mathcal{L}_L,\mathcal{R},\mathcal{L}_R$ standing for either the left lead, the ring or the right lead. In these equations, we have also introduced the following notation
\begin{equation}
H_X = P_X H P_X
\end{equation}
that denotes the sub-Hamiltonian associated with the left ($\mathcal{L}_L$) lead, the right ($\mathcal{L}_R$) lead, or the ring ($\mathcal{R}$) obtained by use of previously introduced projectors. The quantity 
\begin{equation}
\Sigma_X = W_{\mathcal{R}X} G_X W_{X\mathcal{R}}
\end{equation}
refers to the self-energy. The poles of a Green function are given by the eigenenergies of the corresponding Hamiltonian. Self-energies modify these poles in a way that those also obtain an imaginary part. These new poles can be interpreted as decaying resonance states with an energy given by the real part of these poles and a lifetime associated with the imaginary part. We have also introduced the matrix elements at each junction site as
\begin{equation}
W_{X\mathcal{R}} = P_X H P_\mathcal{R} = W_{\mathcal{R}X}^\dagger
\end{equation}
which are explicitly given by
\begin{align}
W_{\mathcal{L}_L\mathcal{R}} = - \dfrac{E_\delta}{2} \ket{-1}\bra{0}, \\
W_{\mathcal{L}_R\mathcal{R}} = - \dfrac{E_\delta}{2} \ket{N_R + 1}\bra{\frac{N_R}{2}}.
\end{align} 

In view of first computing the Green function of the ring, we start from the disorder and interaction free ring Hamiltonian
\begin{align}
H_\mathcal{R} & = \sum_{\alpha=0}^{N_R - 1} E_\delta \Big( \ket{\alpha}\bra{\alpha} - \dfrac{e^{-i \Phi}}{2} \ket{\alpha}\bra{\alpha+1} \nonumber \\
& \hspace{2cm} - \dfrac{e^{i \Phi}}{2} \ket{\alpha+1}\bra{\alpha} \Big), 
\end{align}
with $\ket{0} \equiv \ket{N_R}$, where $\alpha$ is an index labelling the ring sites, $N_R$ denotes the number of sites within the ring, $E_\delta$ is the on-site energy and $\Phi$ is the artificial gauge flux. With the help of this Hamiltonian, the free 1D Green function within the disorder-free ring lattice can be written as
\begin{align}
\bra{\alpha}G_\mathcal{R}\ket{\alpha'} & = \dfrac{1}{N_R} \sum_{n=0}^{N_R - 1} \dfrac{e^{2i\pi (\alpha-\alpha') n/N_R }}{\mu - (1 - \cos(2\pi n/N_R - \Phi))E_\delta + i \epsilon} \label{eq:GR1}\\
& = \dfrac{1}{i E_\delta \sin k \delta } \Big[ e^{i\vert \alpha-\alpha'\vert(k\delta \tilde{\Phi})} \sum_{n=0}^{\infty} e^{inL(k\delta + \tilde{\Phi})} \nonumber \\
& \hspace{1cm} + e^{i(L - \vert \alpha-\alpha'\vert)(k\delta - \tilde{\Phi})} \sum_{n=0}^{\infty} e^{inL(k\delta - \tilde{\Phi})}\Big] \label{eq:GR2},
\end{align}
with $\tilde{\Phi} = \Phi \text{ sign}(\alpha-\alpha')$ and $k\delta = \arccos(1-\mu/E_\delta)$ where the identity of Eqs.~\eqref{eq:GR1} and \eqref{eq:GR2} can be shown through the application of the geometric series and Liouville's theorem. The free 1D Green function can thus be seen as a sum over each path linking two sites. Depending on the value taken by $\alpha$ and $\alpha'$, the path can contain several visits of each junction and many explorations of the ring arms. If one now considers smooth and weak disorder, some exponentials appearing in the above Green function evaluation have to be rewritten as
\begin{equation}
e^{iLk\delta} \rightarrow e^{i(\Phi_u + \Phi_d)} 
\end{equation}
with $\Phi_u$ and $\Phi_d$ the phase due to the disorder in the upper and lower arms. One can express those phases as
\begin{align}
\Phi_u & = \sum_{\alpha=0}^{N_R/2-1} k_\alpha,\\
\Phi_d & = \sum_{\alpha=N_R/2}^{N_R-1} k_\alpha,
\end{align}
where $k_\alpha \delta = \arccos(1-(\mu-V_\alpha)/E_\delta)$. We also have to make the following substitution due to the presence of disorder
\begin{equation}
e^{i\vert \alpha-\alpha'\vert k} \rightarrow \exp\left(i \text{sign} (\alpha-\alpha') \sum_{\alpha'' = \alpha'}^{\alpha-1}k_{\alpha''}\right).
\end{equation} 

If we further consider the fact that the ring in presence of disorder is coupled to the leads, we are left with the calculation of $\left(G_\mathcal{R}^{-1} - (\Sigma_{\mathcal{L}_L} + \Sigma_{\mathcal{L}_R})\right)^{-1}$ that can be expanded in a perturbative series of the self-energies, hence giving rise to a Dyson series
\begin{align}
\left(G_\mathcal{R}^{-1} - (\Sigma_{\mathcal{L}_L} + \Sigma_{\mathcal{L}_R})\right)^{-1} = G_\mathcal{R} + G_\mathcal{R} (\Sigma_{\mathcal{L}_L} + \Sigma_{\mathcal{L}_R}) G_\mathcal{R} \nonumber\\
+ G_\mathcal{R} (\Sigma_L + \Sigma_R) G_\mathcal{R} (\Sigma_{\mathcal{L}_L} + \Sigma_{\mathcal{L}_R}) G_\mathcal{R} + \ldots
\end{align}
The total Green function of the system is then expressed as a sum over paths within ring with the possibility of reflection, transmission or exit each time a junction is visited. This yields the Green function given by Eq.~\eqref{eq/GF_SumSCtraj}.
\vspace{0.5cm}
\section{Techniques and results to calculate further resummed
  diagrams}
\label{sec:appendix_diagrammatic}
By solving the linear diagrammatic equation,
Eq.~(\ref{eq:selfcons_matrix}) we obtained the result
(\ref{eq:LD_Lx_result}). From that it is easy to derive other resummed
diagrams. We obtain the following diagrammatic identities, and calculate
  \begin{equation}
    \begin{aligned}
    &\includegraphics[width=0.6\linewidth]{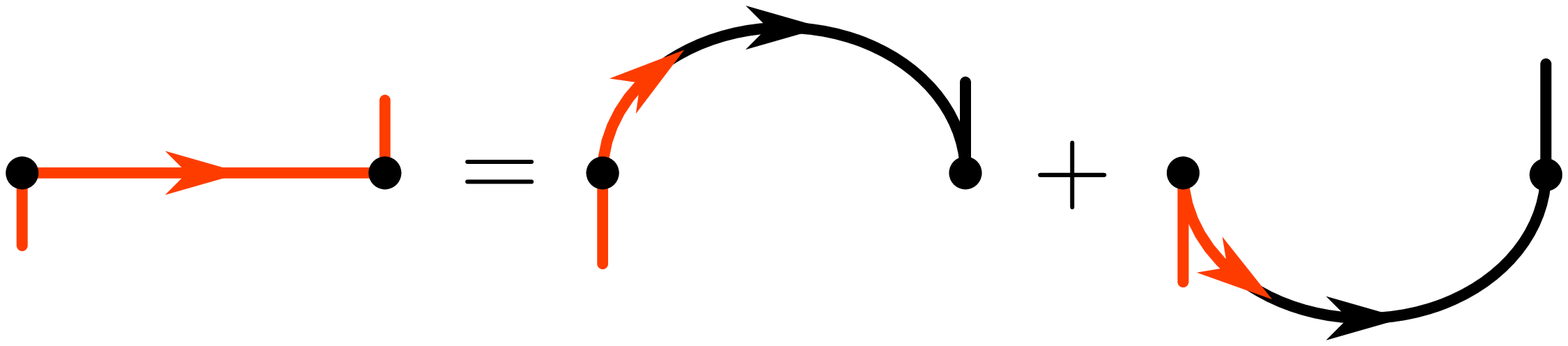}\\
    &=
      e^{-i\left(\pu-\frac{\p}{2}\right)}
      \frac{
      rt
      \left(
      e^{2 i \pu}
      +
      e^{i\left( \pu + \pd -\p\right)}
      \right)}
    {D(\pu,\pd,\p)}\,,
  \end{aligned}
\end{equation}
\begin{widetext}
  \begin{align}
    \includegraphics[width=0.3\linewidth]{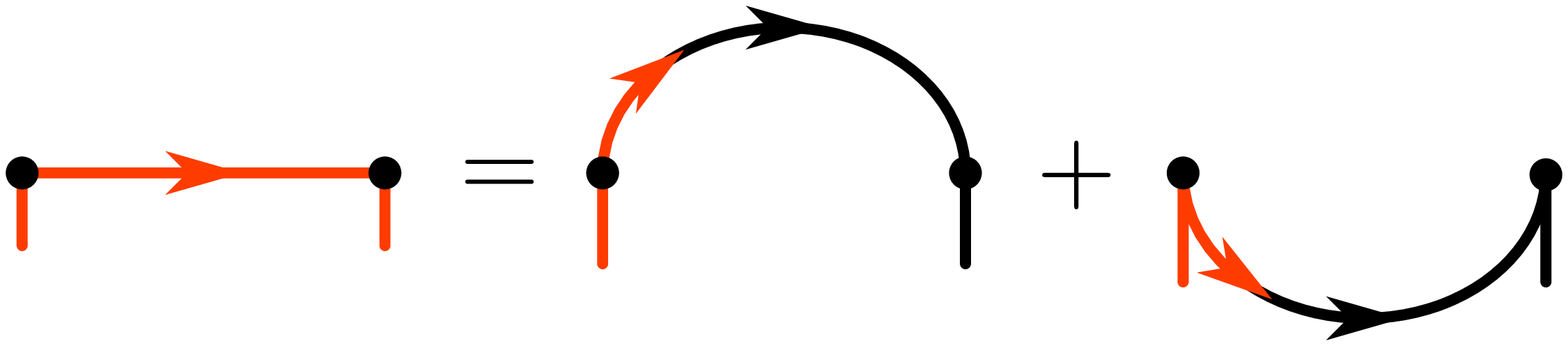}
    & = e^{-i \left( \pd + \frac{\p}{2} \right)}
      \frac{
      r^2e^{2 i \pd }
      +
      t^2e^{i\left(\pu + \pd +\p \right)}
      -\left(r^2-t^2\right)^2
      e^{2i\left(\pu+\pd \right)}
      }{D(\pu,\pd,\p)}
      \,,
    \\
   \includegraphics[width=0.3\linewidth]{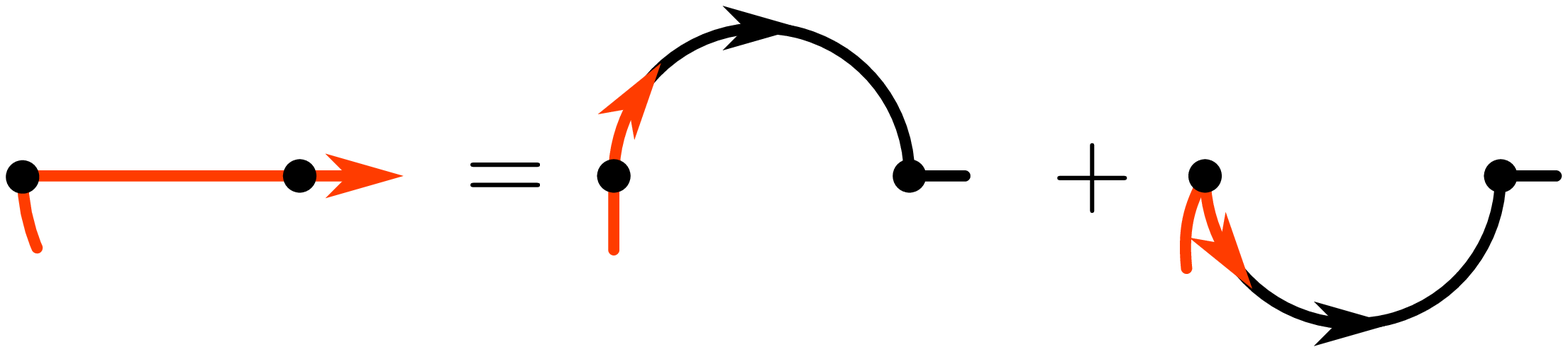}
    &=  e^{-i\left(\pd+\frac{\p}{2}\right)} t
      \frac{r e^{i 2\pd} + t
      e^{i\left(\pu+\pd +\p \right)}
      -\left(r-t\right)\left(r^2-t^2\right)e^{i 2\left(\pu+\pd\right)}
      }
      {D(\pu,\pd,\p)}
      \,,
  \end{align}
\end{widetext}
\begin{equation}
  \begin{aligned}
    &\includegraphics[width=0.6\linewidth]{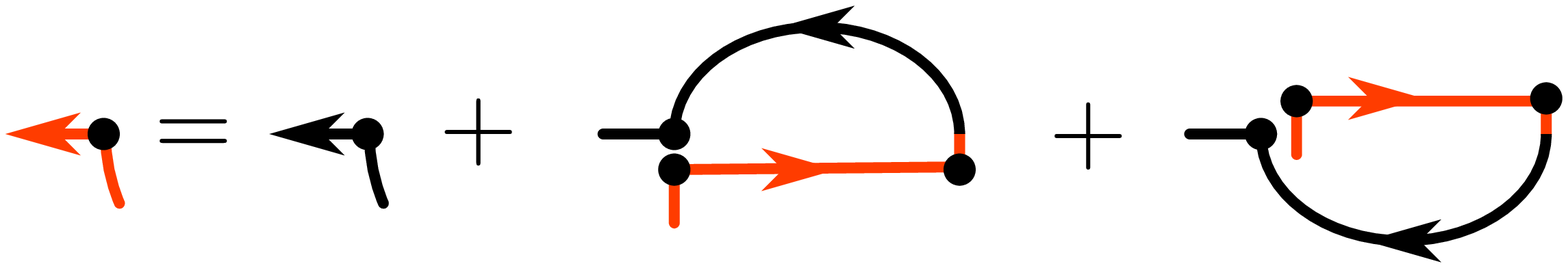}\\
    &  = t
    \frac{1-\left(r-t\right)
      \left(r e^{i 2\pu} - t
        e^{i\left(\pu+\pd +\p \right)}
      \right)
    }
    {D(\pu,\pd,\p)}.
  \end{aligned}
\end{equation}
Mirrored, rotated and time-reversed versions of the already calculated
diagrams can be easily derived from the latter by utilising the
following observation:
\begin{itemize}
\item Inverting the direction of a resummed diagram is done by
  reverting every direction in the trajectories contained in its
  sum. This inversion does not change the branches which are explored,
  but clockwise exploration is transformed into counterclockwise
  exploration. The enclosed flux is thus encircled in opposite
  direction, leading ultimatly to a sign change of the phase $\p$ in the
  results for the diagrams.
\item Mirroring a diagram along the horizontal axis of the ring leads to
  an interchange of the upper and the lower branch of the ring, and
  thus to an interchange $\pu\leftrightarrow\pd$ of the associated
  disorder phases. Since clockwise exploration of the branches turns
  into counterclockwise and vice versa, we also have to change the sign
  of $\p$ again.
\item Mirroring a diagram along the vertical symmetry axis of the ring
  does not change the explored branches, but interchanges clockwise
  and counterclockwise motion, which yet again flips the sign in front of
  $\p$.
\end{itemize}

\bibliography{bibtex/Julien,bibtex/ComScal,bibtex/GaugeFields,bibtex/AtomLaser,bibtex/ring,bibtex/Peter,bibtex/Peierls,bibtex/CBS,bibtex/AB,bibtex/AndLoc,bibtex/TruncWig,bibtex/Transport,bibtex/graphs}

\end{document}